\documentclass[aps,prx,twocolumn,showpacs,superscriptaddress,groupedaddress, nofootinbib]{revtex4} 

\usepackage{graphicx}  % needed for figures
\usepackage{dcolumn}   % needed for some tables
\usepackage{bm}        % for math
\usepackage{amssymb}   % for math
\usepackage{float}

\usepackage{amsmath,mathrsfs}
\usepackage{amsthm}
\usepackage{epsfig}

\usepackage{xcolor}
\usepackage{soul}

\usepackage[colorlinks=true, urlcolor=violet, linkcolor=blue, citecolor=red, linktocpage=true]{hyperref}

\begin{document}

\title{A simple model for multiple-choice collective decision making}

\author{Ching Hua Lee}
\email{clee2@stanford.edu}
\affiliation{Department of Physics, Stanford University, Stanford, CA 94305, USA}

\author{Andrew Lucas}
\email{lucas@fas.harvard.edu}
\affiliation{Department of Physics, Harvard University, Cambridge, MA 02138, USA}

\date{\today}

\begin{abstract}
We describe a simple model of heterogeneous, interacting agents making decisions between $n\ge 2$ discrete choices.   For a special class of interactions, our model is the mean field description of random-field Potts-like models, and is effectively solved by finding the extrema of the average energy $E$ per agent.  In these cases, by studying the propagation of decision changes via avalanches, we argue that macroscopic dynamics is well-captured by a gradient flow along $E$.   We focus on the permutation-symmetric case, where all $n$ choices are (on average) the same, and spontaneous symmetry breaking (SSB) arises purely from cooperative social interactions.  As examples, we show that bimodal heterogeneity naturally provides a mechanism for the spontaneous formation of hierarchies between decisions, and that SSB is a preferred instability to discontinuous phase transitions between two symmetric points.  Beyond the mean field limit, exponentially many stable equilibria emerge when we place this model on a graph of finite mean degree.   We conclude with speculation on decision making with persistent collective oscillations.   Throughout the paper, we emphasize analogies between methods of solution to our model and common intuition from diverse areas of physics, including statistical physics and electromagnetism.
\end{abstract}

\maketitle

\tableofcontents

\section{Introduction}
People  have long imagined connections between social decision making problems and disordered spin systems in physics \cite{brock, Samanidou2007, Castellano2009, Bouchaud2013}.   Many models \cite{Bouchaud2013, Galam1997, Galam1991,  Watts2002, Crucitti2004, Bouchaud2005, Gordon2013, Lucas2013, Lucas2013b, Bouchaud2014} have showed that the most basic binary decision making problems exhibit a variety of interesting behavior such as phase transitions and ``glassy" behavior, with emphasis often placed on the random field Ising model \cite{Bouchaud2013, Galam1997} for its simplicity.   One of the long-term goals of these works is to provide toy models for a variety of social phenomena that are notoriously challenging to explain using the traditional language of economics.   The most basic example is the interpretation of market crashes as discontinuous phase transitions, which naturally arise in spin models placed under external magnetic fields;   the spins represent the agents in the market, and the external magnetic field represents some external ``market forces" that drive these agents towards a specific decision.

However, one aspect of decision making which has been mostly taken for granted is the possibility that there are more than two choices to make.   With few exceptions \cite{Borghesi2006,  Lorenz2009, Davis2013, Salganik2006,Raffaelli2005}, in general this possibility has been assumed to be well approximated by the binary decision making case.   Even in these papers, there is very little analytic development of a theory of $n$-ary decision making, as there has been in the binary decision literature.   A first step towards analytic development of a theory was taken in \cite{Blumm2012}, in a very simple toy model describing the dynamics of a ranking system, although only at the macroscopic level; a case with only ranking preferences considered is given in \cite{Raffaelli2005}. There, an interesting alternative multiple-decision making approach was presented in a rigorous statistical framework.

In this paper, we write down a simple model for decision making between interacting, heterogeneous agents choosing between $n$ possible options.   The agents are forced to pick exactly one option, in contrast with \cite{Borghesi2006}, e.g.   We show that for a wide variety of social interactions, the mean field limit of the model is equivalent to a generalized class of random-field Potts models.   In this case, we will argue that essentially all static and dynamic aspects of the model, at mean field level, can be understood by determining a mean-field energy functional.  This allows us to make strong statements about the resulting phase diagram and dynamics of the model, under a wide variety of random fields and initial conditions.

Our paper is organized as follows:  In Section \ref{sec2}, we introduce our decision model from the viewpoint of local utility maximization, under the assumption that the utilities are dependent on both an ''intrinsic'' component and a ''social'' component dependent on the actions of others. From that we describe a generic framework for understanding equilibria in terms of an energy function $E$.  Section \ref{sec3} re-derives a large class of these models from the microscopic Hamiltonian approach of statistical physics. To make our results more concrete, we present some simple solutions of our model in Section \ref{sect:examples}.   In Section \ref{sec5} we argue that the dynamics of our model is captured by a gradient flow on this energy function $E$, and in Section \ref{SSBpermsymm} we discuss patterns of spontaneous symmetry breaking (SSB), whose general introduction can be found in Appendix \ref{ssbintro}. In Section \ref{sec:complexity} we discuss the emergence of many solutions to the equations of state beyond the mean field limit, and discuss finite size effects in Section \ref{secfinite}.  We conclude with a speculative discussion of a scenario without a globally defined energy that exhibits persistent oscillations in Section \ref{sec8}.

%The goal of this paper is to demonstrate that, in fact, a variety of rich and interesting phenomena are possible in $n$-ary decision models for $n>2$, many of which do not show up for $n=2$. We propose a simple but very general ansatz of cooperative social interactions whose mean-field behavior is elegantly captured by the dynamics on an effective scalar potential. This physical picture allows one to understand certain properties of a market, i.e. its phase transitions (crashes) and stability in terms of the ever-popular theme of symmetry breaking. %It also allows for a general classifcation of markets through mathematical methods like singularity theory and algebraic topology, which is beyond the scope of most other decision models in the literature. 
%Microscopically, our model coincides reduces to a large class of Random-Field Potts Models (cite), thus providing an intuitive solution these models in the mean-field limit. This provides yet another rigorous link between statisical mechanics and decision-making models. 
%
%In this paper, we shall first introduce the our multiple decision model and present a few quick examples. Next, we shall show that our model can indeed be 'derived from' Random-Field Potts Models (RFPM), and ...

%\textbf{elaborate on the merits of such a model.... give some quick examples}
%\textbf{introduce the main themes like SSB, dynamics and phase transitions, cool mathematical methods, complexity etc}

\section{The Model}\label{sec2}

Suppose that we have a collection of $N$ agents, each represented by nodes $\alpha = 1,\ldots, N$ on a graph. Each agent has to make a choice between $n$ possible options, and we call that choice $x_\alpha \in \lbrace 1,\ldots, n\rbrace$. For instance, the agents may represent voters in an election, or consumers deciding between different social media platforms, smartphones or other sets of goods belonging to the same niche.

The key premise in our model is that each agent always selects the choice he deems to have the highest total utility. The total utility is the sum of two components, the intrinsic utility and the social utility. For agent $\alpha$ and choice $i$, we write
\begin{equation}
V_{\alpha,i} = U_{\alpha,i} + f_i(q_1^\alpha,\ldots, q_n^\alpha)%=U_{\alpha,i} + f_i(\vec q^{\;\alpha}),   
\label{eq2}
\end{equation}
where $V_{\alpha,i},U_{\alpha,i}$ and $f_i$ respectively represent the total, intrinsic and social utilities, and $q_i^\alpha$ is the \emph{fraction} of neighbors of node $\alpha$ that subscribe to choice $i$.  As in previously proposed decision-making models like \cite{Watts2002}, the social utility depends only on the \emph{relative} number of agents who prefer some choices over others.  For most of this paper, we will focus on the mean-field limit where the graph is complete, i.e. that $q_i^\alpha$ is independent of $\alpha$. Henceforth, we shall use the vector notation $\vec q = (q_1,...,q_n)$, without the $\alpha$ index.    

Of course, \begin{equation}
\sum_{i=1}^n q_i = 1.  \label{simplexeq}
\end{equation}Along with the constraint that $q_i \ge 0$ for each $i$, this defines a space commonly called the $(n-1)$-dimensional \emph{simplex}.\footnote{This can be thought of as the generalization of the triangle $n=2$, or tetrahedron $n=3$.}   It enjoys a large discrete symmetry group called the permutation group $\mathrm{S}_n$:  the space looks identical if we permute the labels $i$.

The intrinsic utility $U_{\alpha,i}$ encompasses all heterogeneity considerations among agents that do not depend on the choices of other agents, such as price or software reliability.  It is a quenched random variable with cumulative distribution function (CDF) \begin{equation}
F_i(u) \equiv \mathrm{P}(U_{\alpha,i}<u).
\end{equation}
We assume that the agents $\alpha$ always choose the option with the highest total utility:\begin{equation}
x_\alpha = i\text{ if } V_{\alpha,i} > V_{\alpha,j} \text{ for all } i\ne j.
\label{x2}
\end{equation}
  Since we will assume that $V_{\alpha,i}$ are continuous parameters, almost surely we will never have $V_{\alpha,i} = V_{\alpha,j}$ for $i\ne j$, and we neglect this possibility from here on out.   We will also assume for this paper that $V_{\alpha,i}$ and $V_{\alpha,j}$ are uncorrelated for each $i\ne j$.    In practice, this may be a bad assumption, but it will allow us to take advantage of more powerful analytic tools.  There are many other threshold models where agents only change after pushed past some critical threshhold, dependent on the action of others \cite{Watts2002, Crucitti2004, granovetter, valente, llas} -- see also the fiber bundle model \cite{dhkim, pradhan}.    Most of the above works emphasize the possibility that social interaction can alter the phase diagram, with phase transitions characterized by avalanches of macroscopically many state changes;  the model we describe will exhibit such features too, as we shall rigorously derive.
  
So far, we have made no assumptions on the forms of the intrinsic utility CDFs $F_i(u)$, nor those of the social utility functions $f_i$. Indeed, our model is valid for the \emph{most generic} case where these utility functions are nonlinear, and different for each choice. For instance, we can model a situation where the market consists of both normal goods and luxury goods, the latter whose utility \emph{increases} the more expensive and uncommon it becomes.

  We have not included any noise in the model at this point.   As such, we expect this class of models to be a poor description of say, stock trades, which can occur on rapid time scales and are characterized by noisy dynamics.   Instead, we expect this class of models to be better suited for studying decision making which occurs on longer time scales -- say, in the market between two different types of cars, or different neighborhoods to live in. 
  
In the limit where interactions are negligible, this model can be used as a ``microscopic justification" for classical economics, as follows.    Suppose that a good in a marketplace is being sold at price $p$.   For simplicity, let us assume that there are only two choices, and that choice 2 provides no utility:  $F_2(u) = \Theta(u)$.   We are free to choose\footnote{This is because utility is not well-defined:  utility functions $u$ are equivalent to $f(u)$ if $f$ is monotonically increasing ($f^\prime >0$).}
$U_{\alpha 1} = U_{\alpha 0}-p$, where $U_{\alpha 0}$ is the intrinsic utility of the good at price $p=0$.    If $F_1(u)$ is the CDF of $U_{\alpha 0}$, then we conclude that the fraction of buyers (agents in state 1) at price $p$, the demand curve $q_1(p)$, is given by $q_1(p) = F_0(p)$.   The existence of a single-valued demand curve is equivalent to the statement that there is an (effective) description of agents who are non-interacting.

But as we will emphasize repeatedly in this paper, the presence of social interactions generically destroys this picture:  demand becomes a multi-valued function \cite{Gordon2013}.   In general, we can obtain the mean field (MF) equilibria $\vec q$ by self-consistently solving the equations above.  Since $F_i(u) \equiv \mathrm{P}(U_{\alpha,i}<u)$, we have \begin{equation}
\mathrm{P}(V_{\alpha,i}<u) =\mathrm{P}(U_{\alpha,i}<u-f_i)= F_i(u-f_i)
\end{equation}
Hence the MF equilibria can be obtained by integrating over all possible values $u$ that the most desirable choice could be, multiplied by the probability that that choice is $i$ \emph{and} the value of all other choices is smaller:
% \begin{equation}
%q_i = \mathrm{P}(x_\alpha=i)=\int\limits \mathrm{d}u F_i^\prime(u-f_i(\vec q)) \prod_{j\ne i} F_j(u-f_j(\vec q)).   \label{mfeq}
%\end{equation}
\begin{eqnarray}
q_i &=&\mathrm{P}(x_\alpha=i)\notag\\
&=&\int\limits \mathrm{P}(u<V_{\alpha,i}<u+\mathrm{d}u)\prod_{j\ne i}\mathrm{P}(V_{\alpha,j}<u)\notag\\
&=&\int\limits \mathrm{d}u \;  F_i^\prime(u-f_i(\vec q)) \prod_{j\ne i} F_j(u-f_j(\vec q))\notag\\
&=& -\frac{\partial}{\partial f_i}\int\limits_{-\infty}^\infty \mathrm{d}u \prod_{i=1}^n F_i(u-f_i(\vec q))
\label{mfeq}
\end{eqnarray}
The MF approximation becomes exact as $N\rightarrow \infty$.

Note that an overall uniform shift in the value of \emph{each} $f_i$:   $f_i \rightarrow f_i + a$, does not change our model; only relative social utilities matter.  The MF equilibrium $\vec q$ can be written in terms of a potential\footnote{The integral given by Eq. (\ref{geq}) is formally infinite, as $\prod_i F_i \rightarrow 1$ for large $u$, but this infinity can be trivially regulated by replacing the upper bound in the integral above by $R$, and taking the limit $R\rightarrow\infty$.   As only derivatives of $G$ enter the equation for $\vec q$, the overall linear coefficient $R$ in $G$ will be irrelevant for physical calculations.  }
\begin{equation}
G(\vec q) \equiv \int\limits_{-\infty}^\infty \mathrm{d}u \prod_{i=1}^n F_i(u-f_i(\vec q)),   \label{geq}
\end{equation} via \begin{equation}
q_ i =-\frac{\partial G}{\partial f_i}.
\label{eqm1}
\end{equation}  
For generic nonlinear functions $\vec f(\vec q)$, the derivative $\partial/\partial f_i$ may not be globally defined, as we discuss in Appendix \ref{appgradflow}.    

%In the following section, we shall provide a microscopic description of our decision-making model based on a class of statistical models that is ubiquitous in physics. Readers who are interested in concrete decision-making applications may skip straight to section \ref{sect:examples}.

\section{The Hamiltonian Approach}\label{sec3}
Now, we (a priori) start from a very different perspective, motivated from statistical physics.   Let us imagine that there is some global function:  the Hamiltonian $H$, which must be a local extremum at a stationary point of the dynamics.    For a social system, this is an unjustified assumption.   Nonetheless, we will show that we can recover a large class of $\vec f(\vec q)$ from this approach, so in retrospect it may be reasonable, and so we proceed.

A microscopic Hamiltonian that can describe social decision making is
\begin{equation}
H = -\sum_i U_{i\alpha} z_{i\alpha} - \sum_{n\ge 2,i_1,\ldots,i_n} \frac{A^{(n)}_{i_1\cdots i_n}}{nN^{n-1}} z_{i_1\alpha_1}\cdots z_{i_n\alpha_n}.
\end{equation}
where the indices $\alpha$ and $i$ refer to the nodes and their internal states respectively. $z_{i\alpha} =0$ or 1, and satisfy the constraint \begin{equation}
\sum_{i=1}^n z_{i\alpha}=1.
\end{equation}
We interpret $z_{i\alpha}=1$ as the state where node $\alpha$ is in state $i$.   The $U_{i\alpha}$'s, which are 'single particle' energies associated with node $\alpha$ being in state $i$, are random variables which are drawn from the cumulative distribution function $F_i(u)$, as in the previous section.   The notation has been chosen identically with the previous section because we will shortly show that the $U_{i\alpha}$s are playing an identical role.   For the thermodynamic limit to be well-defined, we require that $A^{(n)}$ are independent of $N$.

In the case where the only non-vanishing $A^{(n)}$ is $A^{(2)}$, and $A^{(2)}_{ij} = A\delta_{ij}$, the Hamiltonian we have written down is simply the random field Potts model \cite{nishimori} on a complete graph.   The statement that our model lives on a complete graph is, for introductory purposes, simply a complicated way of saying that there is a contribution to $H$ from every single pair of agents.    Later in the paper, we will discuss physics on more complicated graphs, where not all agents interacts with each other.

We can find local extrema of $H$ by demanding that we find a solution where there is no single agent $\alpha$ who can lower $H$ by changing the $i$ for which $z_{i\alpha}=1$.   A sufficient condition for this is that \begin{equation}
z_{i\alpha}=1 \iff -\frac{\partial H}{\partial z_{j\alpha}}\text{ maximal for }j=i.  \label{zia1}
\end{equation}If we define, employing a summation convention on the $i$'s,\begin{equation}
\mathcal{H} \equiv -\sum_{n\ge 2} \frac{A^{(n)}_{i_1\cdots i_n}}{n} q_{i_1}\cdots q_{i_n},
\end{equation} \begin{equation}
f_i \equiv -\frac{\partial \mathcal{H}}{\partial q_i} =  \sum_{n\ge 2} A^{(n)}_{ij_2\cdots j_n} q_{j_2}\cdots q_{j_n},
\end{equation}then noting that the sum over $z_{i\alpha}/N$ over all agents $\alpha$ tends to $q_i$ in the thermodynamic limit, we obtain that Eq. (\ref{zia1}) is equivalent to 
\begin{equation}
z_{i\alpha}=1 \iff U_{j\alpha} + f_j \text{ maximal for } j=i.  \label{zia2}
\end{equation}This precisely corresponds to our defining equation for the model in the previous section, which we derived by an analogy to utility.     

From the above equations, it is clear that a positive (negative) $A^{(n)_{i_1\cdots i_n}}$ determines whether the system prefers (does not prefer) the choices represented by $q_{i_1}$,...,$q_{i_n}$ to be realized simultaneously. For instance, a negative $A^{(2)}_{i_1i_2}$ implies that a higher market share for one choice discourages the other choice from being taken up. In the case of equal $i_1$ and $i_2$, the sign of $A^{(2)}$ simply determines whether the social utility of the choice is positive or negative. 

We will essentially restrict our analysis to the case where all $A^{(n)}_{i_1\cdots i_n} >0$, and will comment on the opposite case in Appendix \ref{appgradflow}.

\subsection{An Energy Function}
There is a remarkably easy and intuitive way to find the solutions to the mean field equation Eq. (\ref{zia2}), given that $\vec f = -\nabla_q \mathcal{H}$.   To see this, let us recall Eq. (\ref{eqm1}), which states that $\vec q = -\nabla_f G$, for a scalar function $G$.   Assuming that the matrix $\partial f_i / \partial q_j$ is invertible everywhere, we can re-write Eq. (\ref{eqm1}) as \begin{equation}
0 = q_i \frac{\partial f_i}{\partial q_j} + \frac{\partial G}{\partial q_j} = \frac{\partial}{\partial q_j} \left(G + q_i f_i\right) - \delta_{ij}f_i.   \label{eq14}
\end{equation} Using that $\vec f$ itself is a gradient, we see that the solutions to our mean field equations precisely correspond to the extrema of an ``energy" function \begin{equation}
E \equiv G + \mathcal{H} + q_i f_i = G+\mathcal{H} - q_i \frac{\partial \mathcal{H}}{\partial q_i}. 
\end{equation}

In the subsequent subsection, we shall see that $E$ can be physically interpreted as the microscopic Hamiltonian $H$ averaged over disorder:
\begin{equation}
E = \frac{\langle H\rangle}{N},  \label{eqehn}
\end{equation}
up to a possible constant shift, with the averages taken over the random intrinsic utilities $U_{i\alpha}$.   %We will see in the  an elegant way to derive this equation.   

It is tempting to identify the maxima of $E$ with unstable mean field solutions, and the minima of $E$ with stable solutions.  We will show later that, up to a subtlety associated with cooperative (ferromagnetic) vs. antagonistic (antiferromagnetic) interactions, this is indeed the case.  Furthermore, the dynamics will always drive the system towards stable fixed points.

\subsection{Noise}
Noise is an unavoidable feature of any realistic social decision making process.  Noise can take on a variety of forms:  individual uncertainty, noisy stock market dynamics, etc.   It is important to stress, however, that noise is qualitatively different than the random field disorder $U_{i\alpha}$ -- unlike $U_{i\alpha}$, the presence of noise will tend to drive people between different states over time.   

Since we have a Hamiltonian framework for our model, there is a natural way that we can state our ignorance about the true state of the system, and of the noise driving it, by solving our model at finite temperature $T$.    Mathematically, this is a statement that we wish to find the maximal entropy distribution consistent with knowledge of the typical value of $H$, and thus remain ``as uncertain as possible" \cite{jaynes}.

As usual, at finite temperature $T$, the probability of being in any given state is proportional to $\mathrm{e}^{-H/T}$.   Since the change in the energy due to the change of agent $\alpha$ into state $i$ is given by $U_{i\alpha}+f_i$, the probability that a node with given $U_i$ will be in state $i$ is given at mean field level by \begin{equation}
\mathrm{P}(z_{i\alpha}=1|U_{j\alpha}) = \frac{\mathrm{e}^{(U_i+f_i)/T}}{\sum \mathrm{e}^{(U_j+f_j)/T}}.
\end{equation}Thus, $q_i$ is given by simply averaging the above equation over disorder.    In particular, we write \begin{equation}
q_i = -\frac{\partial G_T}{\partial f_i},
\end{equation}where \begin{equation}
G_T \equiv -T \left\langle \log \left(\sum \mathrm{e}^{(U_i+f_i)/T}\right) \right\rangle_U
\end{equation}with $\langle \cdots \rangle_U$ denoting disorder (but not thermal) averages.  

The analogy with the function $G$ defined previously is not accidental: \begin{equation}
\lim_{T\rightarrow 0} G_T = -\langle \max_i (U_i + f_i)\rangle_U.
\end{equation}Taking the derivative of this with respect to $f_i$, we find that this is simply equal to the probability that $U_i + f_i$ is maximal, which is precisely the $f_i$ derivative of our previous $G$ function.    Therefore, up to a (infinite) constant, $G=G_0$.    It is now easy to derive Eq. (\ref{eqehn}): \begin{align}
\frac{\langle H\rangle}{N} &= \mathcal{H} - \left\langle \sum U_{i\alpha} \frac{z_{i\alpha}}{N}\right\rangle \notag \\
&= \mathcal{H} - \sum q_i \langle U_i | U_i + f_i \text{ maximal}\rangle \notag \\
&= \mathcal{H} + q_i f_i  - \langle \max (U_i + f_i) \rangle \notag \\
&= \mathcal{H}+G_0 + q_if_i.
\end{align}We conclude that $G_0 + q_i f_i$ can be interpreted as the random-field-induced potential energy in the effective, disorder-averaged Hamiltonian (at mean field level).

By replacing $G$ with $G_T$, we obtain the free energy per spin, which should be an extremum in equilibrium at finite temperature.

\section{Simple Examples}
\label{sect:examples}

\subsection{Binary decision models ($n=2$)}

We first consider the case where each agent decides between $n=2$ options, i.e. between two difficult products, or whether to buy or not to buy a single product. For simplicity, we shall use the latter interpretation below.    These models are extensively studied \cite{Bouchaud2013, Galam1997, Galam1991,  Watts2002, Crucitti2004, Bouchaud2005, Gordon2013, Lucas2013, Lucas2013b, Bouchaud2014}.

This potential description also leads to a simpler understanding of a similar binary-decision model discussed in \cite{Lucas2013}.  It enables one to graphically read off the stability of a fixed point, and more importantly, generalizes very easily to the case $n>2$, as we will discuss in the next subsection.

\subsubsection{Homogeneous intrinsic utility}

In this simplest scenario, each agent have exactly the same, i.e. homogeneous intrinsic opinion on the value of the product. Each buyer will be rewarded with an intrinsic utility of $U_0$, and each non-buyer with zero utility. This corresponds to CDFs
\begin{equation}
F_1(u)=\Theta(u-U_0);\;\;\;F_2(u)=\Theta(u)
\label{F12}
\end{equation}
where $\Theta(u)$ is the Heaviside function \begin{equation}
\Theta(u) \equiv \left\lbrace\begin{array}{ll} 1 &\ u\ge 0 \\ 0 &\ u<0 \end{array}\right..
\end{equation} Now, also suppose that $\vec f = \vec q$, i.e. each agent ascribes a social utility to each choice proportionally to the fraction of friends already subscribed to it. With $\vec q = (q_1,q_2)=(q,1-q)$, the options of buying/not buying will have total utilities of 
\begin{equation}
V_1 = U_0 + q;\;\;\;  V_2 = 1-q 
\end{equation}

%\begin{table}[H]
%\centering
%\renewcommand{\arraystretch}{2}
%\begin{tabular}{|l|l|l|}\hline
%\textbf{Quantity} &\ \textbf{Choice $1$}  &\ \textbf{Choice $2$}  \\ 
% &\ (buy)  &\ (not buy) \\ \hline
%Intrinsic Utility CDF $F_i(u)$  &\ $\theta(u-U_0)$ &\ $\theta(u) $\\ \hline
%Social Utility $f_i(q)$  &\ $q$ &\ $1-q$ \\ \hline
%%Total Utility $V_i$  &\ $U+q$ &\ $1-q$ \\ \hline
%\end{tabular}
%\caption{The social utilities and the cumulative distribution functions of the intrinsic utilities for each choice. This is the limit of zero 'disorder', when each agent ascribes exactly the same intrinsic utility $U_0$ to the product.}
%\label{table1}
%\end{table}

We are now ready to calculate the potential $G(q)$. Substituting Eqs. (\ref{F12}) into Eq. (\ref{mfeq}), we obtain

\begin{eqnarray}
G(q)&=&\int \mathrm{d}u\; F_1(u-q)F_2(u-(1-q))\notag\\
&=&\int \mathrm{d}u\; \Theta(u-(q+U_0))\Theta(u-(1-q))\notag\\
&=&-\max(q+U_0,1-q)+\text{constant}.
\end{eqnarray} 
The energy is 
\begin{eqnarray}
E(q)&=&G(q)+\mathcal{H}(q)-\sum_{i=1}^2q_i\frac{\partial \mathcal{H}(q)}{\partial q_i}\notag\\
&=&-\max(q-U_0,1-q)+ \frac{q^2+(1-q)^2}{2}.
\end{eqnarray}

%\begin{figure}
%\includegraphics[scale=0.7]{graph1.jpg}
%\caption{(Color online.)  Plots of $G(q)$ and $E(q)$. \red{Will replot this properly}}
%\label{graph1}
%\end{figure}

By inspection (or from Fig. \ref{phase2pot}), the local minima of $E(q)$ lie at $q=0$ for $U_0<-1$, at $q=1$ for $U_0>1$, and at both $q=0$ and $q=1$ for $|U_0|<1$. 

This \emph{all-or-none} behavior is easy to explain: Since every agent has exactly the same intrinsic preferences and are exposed to the same $q$, the fraction of friends buying, they will definitely gravitate towards the same optimal choice. For sufficiently small $|U_0|$, the social utility dominates and \emph{both} $q=0$ and $q=1$ states are actually local optima.   This \emph{bistability} implies that if $U_0$ is time-dependent, we can obtain hysteresis -- which of the two equilibria we are at depends on the previous behavior of $U_0$.

%Suppose there is no intrinsic value in buying the product. Then $U_0=0$, and of course there should be no reason why more agents should buy or not buy, i.e. $q_1=q_2=q=\frac{1}{2}$. Increasing/decreasing $U_0$ should increase/decrease the attractiveness of the product and hence its market share. When $U_0 = \pm 1$, the agents will either all buy or not buy

%This simplest model is unfortunately too rudimentary to allow for mean-field optima at arbitrary $q$, as in almost all real-world situations. To model the latter, we have to introduce heterogeneity in the intrinisic utilities, as shown next.

\subsubsection{Variable social and intrinsic utilities}
\label{binarysubsub}

Here, we consider a more general scenario with $n=2$ choices by introducing a spread to the intrinsic utility distribution.  We assume that the utility distribution is unimodal -- i.e. $F^\prime(u)$ has a single maximum.   Intuitively, we imagine all agents have an identical utility $U_0$ (as before) for choice 1, up to random noise which is equally likely to increase or decrease utility.  For qualitative purposes, it suffices to replace the $\Theta$ function in Eq. \ref{F12} with a logistic (Fermi-Dirac) distribution: 
\begin{equation}
F^\prime(u)=\frac{\beta}{4}\mathrm{sech}^2\frac{\beta u}{2}.
\end{equation}
%\begin{equation} P(U=u)=\frac{\beta}{4}sech^2\frac{\beta(u-a)}{2} \end{equation}
This distribution has a variance of $(\pi/3\beta)^2$; its CDF is
\begin{equation} F(u)=\frac{1}{1+\mathrm{e}^{-\beta u}}\label{fd}\end{equation} 
and is useful for analytical studies because an exact expression exists for its corresponding potential $G(\vec q)$, for any $n$ (See Appendix \ref{app:logistic}). Its ``effective temperature" $T\equiv \beta^{-1}$ plays the role of heterogeneity in intrinsic utilities, but we stress that the disorder is \emph{not} thermal in nature despite the suggestive notation -- disorder is quenched, and all agents have the same utility for all time.

Also, we stress that while we will use intrinsic utilities of the form Eq. \ref{fd} extensively in this work, our model is applicable to \emph{all} possible forms of the utility function, including those with an arbitrary number of peaks, or asymmetrical ones.

We may also generalize the social utilities $f_i$ to the linear form
\begin{equation}
 f_i=b_iq_i+a_i 
 \end{equation}
where $b_i$ represents the strength of the social influence of choice $i$, and $a_i$ a mean offset which can also be absorbed into the intrinsic utility.

%\begin{table}[H]
%\centering
%\renewcommand{\arraystretch}{2}
%\begin{tabular}{|l|l|l|}\hline
%\textbf{Quantity} &\ \textbf{Choice $1$}  &\ \textbf{Choice $2$}  \\ \hline
% Intrinsic Utility CDF $F_i(u)$  &\ $\frac{1}{1+e^{-\beta (u-a_1)}}$ &\ $\frac{1}{1+e^{-\beta (u-a_2)}}$\\ \hline
% Social Utility $f_i(q)$  &\ $b_1q$ &\ $b_2(1-q)$ \\ \hline
%%Total Utility $V_i$  &\ $U+q$ &\ $1-q$ \\ \hline
%\end{tabular}
%\caption{The social utilities and the cumulative distribution functions of the intrinsic utilities for each choice....}
%\label{table2}
%\end{table}

The potential $G(\vec f(q))$ can be obtained: 
\begin{eqnarray}
G&=& -\frac{ f_1  \mathrm{e}^{\beta   f_1 }- f_2\mathrm{e}^{\beta   f_2}}{\mathrm{e}^{\beta   f_1 }-\mathrm{e}^{\beta   f_2}}%\notag\\
%&=&  f_2 + \frac{ f_2 - f_1}{e^{\beta( f_2- f_1)}-1}
%&=&\frac{ f_1 + f_2}{2}+\frac{ f_2- f_1 }{2}\coth \frac{\beta ( f_2- f_1 )}{2}\notag\\
%&\sim&\frac{ f_2- f_1 }{2}\coth \frac{\beta ( f_2- f_1 )}{2}
\end{eqnarray}
as derived in Appendix \ref{app:logistic}.  The energy $E(q)$ has a remarkably simple expression:
\begin{eqnarray}
E(q)%&=&G(q)+H(q)-\sum_{q_i}q_i\frac{\partial H(q)}{\partial q_i}|_{q_1+q_2=1}\notag\\
&=&G(q)+ \frac{b_1q^2+b_2(1-q)^2}{2}\notag\\
&=& B\left(\frac{q^2}{2}+\frac{q-C}{\mathrm{e}^{\beta'(q-C)}-1}\right)
\label{E2}
\end{eqnarray}
where $C=A/B$ and $\beta'=\beta B$, with $B=b_1+b_2$ being the total marginal social utility and $A=b_2+a_2-a_1$ the effective intrinsic utility of choice $2$ (perhaps representing the cost of choice $1$). Evidently, the fixed points at $\nabla_q E=0$ only depends on the two external degrees of freedom $C$ and $\beta'$, representing intrinsic imbalances and effective market homogeneity respectively. In particular, $\beta$, which is an intrinsic property of the agents, and $B$, which characterizes their social interactions, play indistinguishable roles in determining the \emph{positions} of market equilibria.    In particular, we will use the ability to absorb the overall strength of social interactions into a rescaling of utilities frequently for the remainder of the paper, and this will henceforth be done implicitly.
%However, the \emph{rate} $\frac{dq}{dt}=-\frac{dE}{dq}$ at which the equilibria are reached depends on $B$ but not $\beta$. 
  
%The potential $E(q)$ leads to a wealth of interesting physical interpretations which are studied in detail in Appendix \ref{choice2}. Here, we shall just focus on a few salient conclusions demonstrated in Fig. \ref{phase2pot}.  

%This solution can be elegantly analysed through graphical means, as shown in Appendix \ref{app:graphical2}, but here we shall focus on the shape of the potential $E$ in the spirit of this section. 

Since $E\sim q^2$ for large\footnote{Note, however, that physically relevant solutions lie within $q\in [0,1]$} $q$, there exists either one unique minimum or two minima and one (unstable) maximum in between them. In the former case, the market will simply gravitate towards the potential minimum. In the latter case, the market will choose the minimum that exists in the same basin of attraction.

\begin{figure}
\includegraphics[scale=0.67]{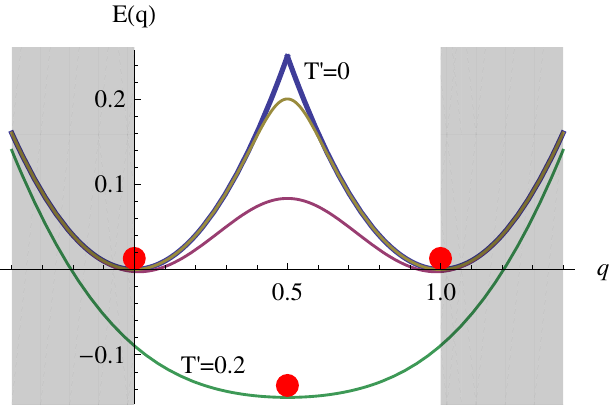}
\includegraphics[scale=0.67]{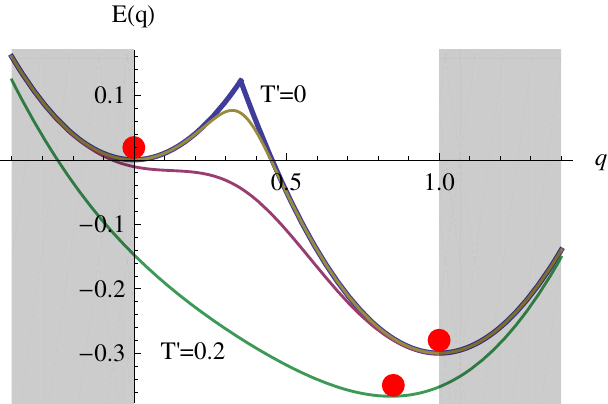}
%ANY EASY WAY TO LEAVE A BIGGER GAP BETWEEN THE TOP AND BOTTOM FIGURES?
\includegraphics[scale=0.693]{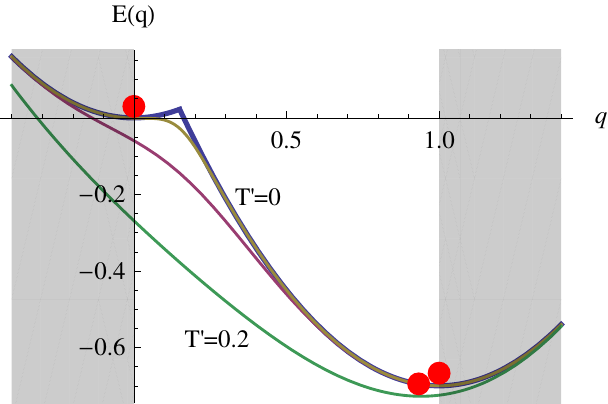}
\includegraphics[scale=0.69]{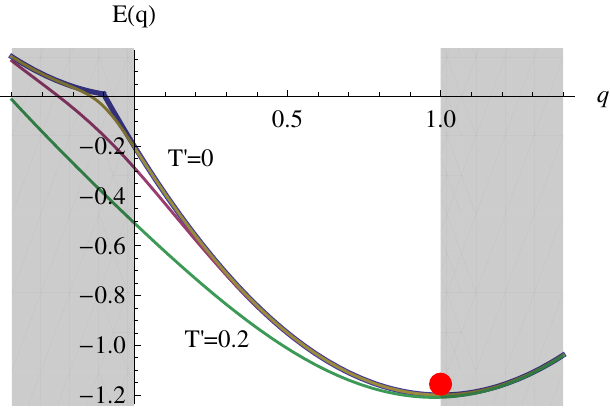}

\caption{(Color Online) Plots of $E(q)$ against $q$ for relative intrinsic utilities $U_0$ corresponding to (a to d) $0, 0.3,0.7,1.2$ (cases with negative $U_0$ are related by mirror symmetry). Here $a_1-a_2=U_0$, and $b_1=b_2=1$. Appearing in each plot are the graphs with $T'=\frac{1}{\beta'}=0,0.025,0.08$ and $0.2$. The red balls indicate local minima in the physically allowed white region. As $U_0$ is increased, the minima with larger $q$ (towards the right) become more favored, until only one minimum eventually remains. 
In general, a larger $T'\propto \sigma$ smooths out potential barriers, thereby undermining social effects. Only when $T'<\frac{1}{6}$ does bistability become a possibility. The curves for $T=0$ reduces to those describing the homogeneous case in previous subsection, where an all-or-none situation prevails.  }
\label{phase2pot}\end{figure}

In Fig. \ref{phase2pot}, we observe how the potential landscape changes for different $T'=\frac{1}{\beta'}$ as $U_0$ is increased from $0$ to $1.2$. As $U_0$ increases, choice $1$ become intrinsically more favorable, and the energy landscape tilts in its favor accordingly. The onset of bistability and hysteresis is set by the competition between quenched disorder $T$ and social forces $B$ via $\beta'=B/T$. While market heterogeneity (quenched disorder) tend to smooth out the potential landscape, social forces encourage 'market crashes', where one local minimum disappears and the state is forced to 'roll down' to another nearest one. The time it takes for the crash to occur decreases as $E(q)$ becomes steeper, and will be examined in detail in Section \ref{sec5} and Appendix \ref{appgradflow}.

 It is particularly important to pay attention to the permutation symmetric case where $a_2=a_1$, and $b_2=b_1$. In this case, any deviation from the point $q=1/2$ represents spontaneous symmetry breaking (SSB), where the agents collectively pick out one choice over the other, even though both choices have the same intrinsic and social utility functions. The phenomenon of SSB will be further elaborated in Section \ref{permsymmcase} on permutation symmetric models.
 
\subsubsection{Nonlinear Logarithmic Utility Function }

\begin{figure}
\begin{minipage}{\linewidth}
\includegraphics[width=.46\linewidth]{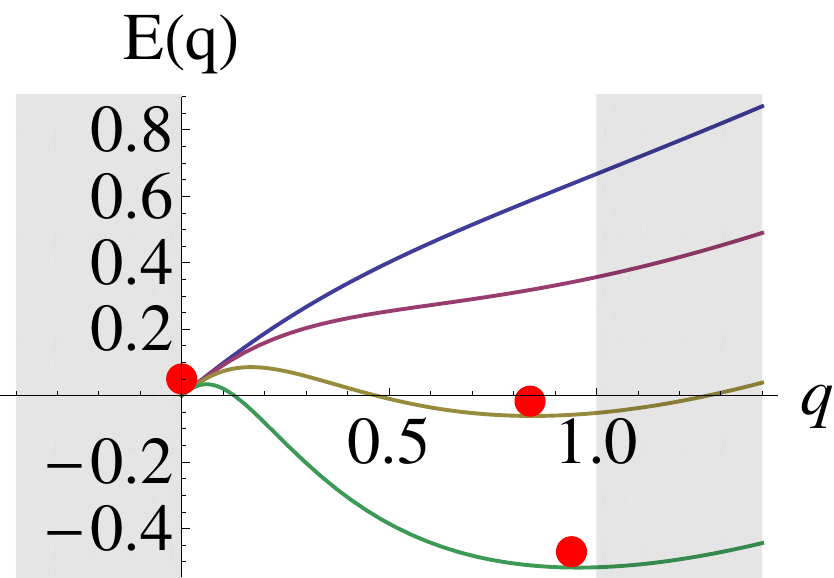}
\includegraphics[width=.46\linewidth]{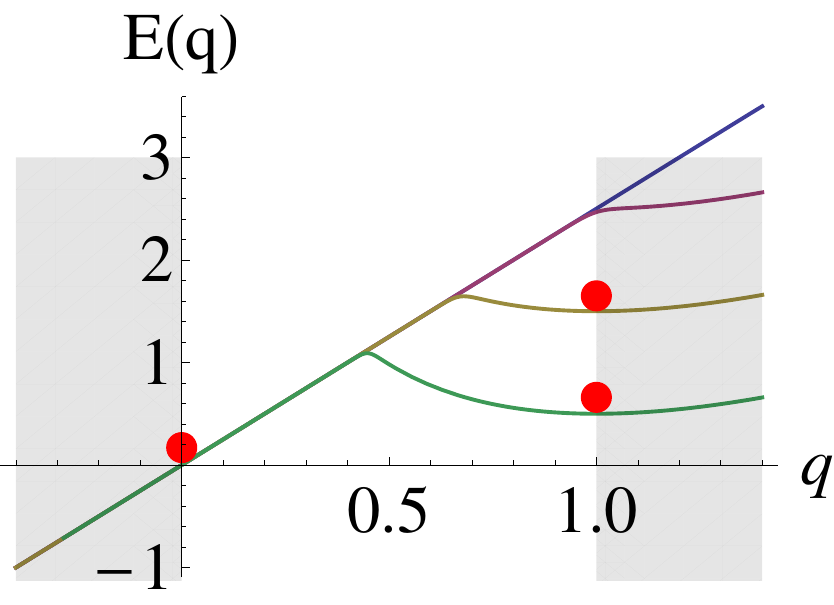}
\end{minipage}
\caption{(Color Online) The potential $E(q)$ for the case of a single logarithimic utility with $\delta=10^{-4}$, with stable MF solutions indicated by red balls. Left) $g_1  =1$, $\beta=3$ and $U_0=0,-0.5,-1,-1.5$ for top to bottom, and Right) $g_1  =2.5$, $\beta=30$ and $U_0=0,-1,-2,-3$ from top to bottom. In both scenarios, we see that the $q\rightarrow 0$ limit is always a local minimum; moreover, it is robust against the tunings of parameters. This is a consequence of the extreme cost of having $q=0$ market share. Also, there is a local maximum when $f_1=f_2$, i.e. $g_1   \log q =U_0$, although it can smoothed out when $T^{-1}=\beta$ is small. It demarcates two basins of attraction, where either choice becomes more attractive. For large $g_1  $ or $\beta$, the social utility dominates, and we have an all-or-none ($q=0$ or $1$) scenario.     }
\label{fig:nonlinearlog}
\end{figure}

Here we briefly consider the case of a nonlinear utility function 
\begin{equation}
f_1=g_1  \log (q+\delta), \;\;\;\; f_2=U_0 
\end{equation}
with $g_1$ a constant.  Logarithmic utility functions feature are commonly used to represent situations where it a deficiency costs much more than it pays to have a surplus.  In particular, they are frequently used in economics because they exhibit ``diminshing returns"  \cite{varian}: $\mathrm{d}f_1/\mathrm{d}q > 0$, but $\mathrm{d}^2 f_1 / \mathrm{d}q^2 < 0$.  $U_0$ represents the fixed utility offset of not buying the product.   Let us suppose that $\delta \ll 1$.  As detailed in Appendix \ref{app:nonlinearlog}, the contribution from $\delta$ can be neglected for small $\delta$, and we have 
\begin{equation}  
E(q)\approx g_1   q + \frac{g_1   \log q - U_0}{\left(q^{g_1}   \mathrm{e}^{-U_0}\right)^\beta-1},  
\end{equation} 
which is graphically analyzed in Fig. \ref{fig:nonlinearlog}.

\subsection{Ternary ($n=3$) Decision Making}
In this subsection we will consider examples of our model with $n=3$.   Unlike the binary case, this case has not received much attention in the literature.     For simplicity, we will assume the social utilities $\vec f = \vec q$ for the remainder of this section.  Details on the case with more general $\vec f(\vec q)$ may be found in Appendix    \ref{app:permsymm}.

\subsubsection{Permutation Symmetric Case}
\label{permsymmcase}
In the permutation symmetric case,
\begin{equation}
F_1(u)=F_2(u)=F_3(u) = F(u),
\end{equation}
so that the intrinsic and social utilities for all choices are identically equal. Any deviation from the permutation symmetric point $q=1/3$ will thus represent spontaneous symmetry breaking (SSB), where the agents collectively pick out one choice over the others. SSB is an extremely important aspect of any model of the decision making process between interacting agents, especially with an eye towards financial or economic applications.   This is because this represents a phenomenon entirely beyond the classical theory of supply and demand.  In particular, SSB implies that a market with identical sellers selling identical goods could still lead to a distorted marketplace where sellers do not receive revenues in accordance with the quality of their product, one of the most foundational principles of competitive market theory.  This point has also been emphasized, e.g., in \cite{Bouchaud2013, Salganik2006}.  See Appendix \ref{ssbintro} for an introduction to SSB and its consequences in physics.

\begin{figure*}
\includegraphics[width=7in]{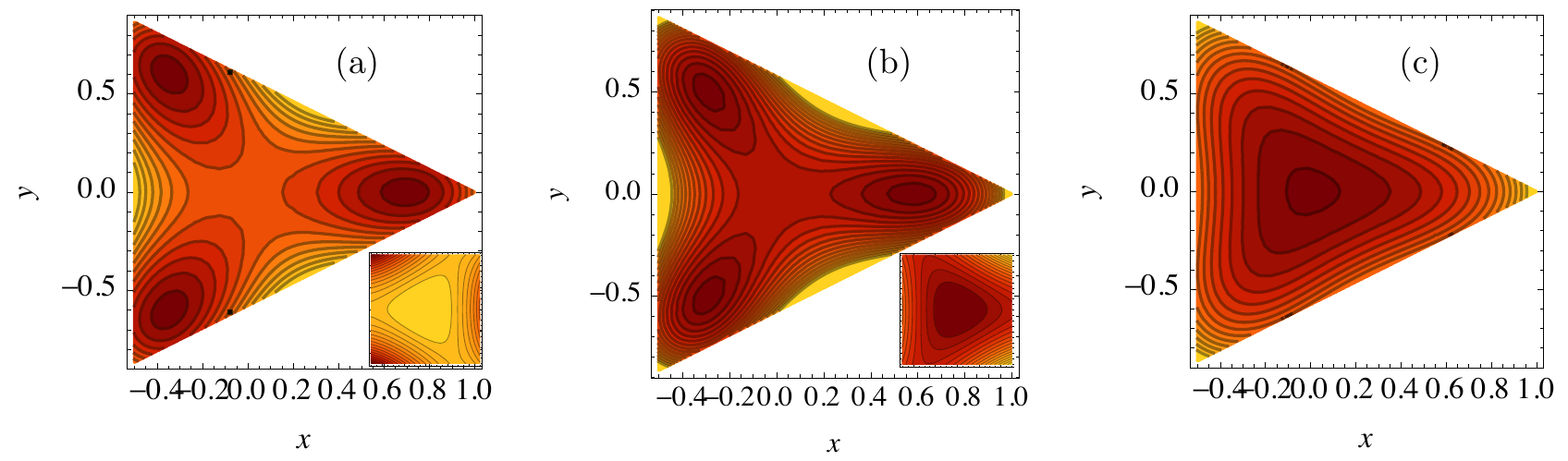}
\caption{(Color online.)  We show the energy landscape $E(x,y)$ for permutation symmetric ternary logistic $F(u)$ with (a) $\beta=4.1$, (b) $\beta=3.9$, (c) $\beta=3.6$. Inset regions in the bottom right correspond to zooming in very close to the permutation symmetric point.   Darker shading corresponds to $E$ smaller, and lighter shading to $E$ larger. As $\beta$ decreases below $4$, the three minima merge to form one permutation symmetric minimum at the center.}
\label{4bfig1}
\end{figure*}

We show the results in Figure \ref{4bfig1}.  For easy plotting of our results in a manifestly permutation symmetric  manner, we have used barycentric coordinates for the simplex, defined by $(x,y)$: \begin{equation}
x+\mathrm{i}y = q_1 + q_2 \mathrm{e}^{2\pi\mathrm{i}/3} + q_3\mathrm{e}^{4\pi\mathrm{i}/3}.   \label{baryeq}
\end{equation}
The simplex $1=q_1+q_2+q_3$ corresponds to the region in between the three lines $x=-1/2$ and $(x-1) =\pm\sqrt{3}y$.

If $F(u)$ is given by a logistic distribution,  by plotting $E$ on the simplex, we numerically find a transition between three different regimes.   From Eq. (\ref{eqd2}) and Appendix \ref{app:permsymm}, for $\beta>4$, the only minima of $E$ correspond to permutation symmetry broken points.   However, for $\beta$ just smaller than 4, we find that although the permutation symmetric point is stable, \emph{new minima} arise in $E$ which break permutation symmetry.  In particular, we find (for example) that $q_1>q_2=q_3$;  there are 3 equivalent points corresponding to which of the three choices is most popular.   Analogously to the ferromagnetic phase of the Ising model, these should be thought of as the \emph{same phase} -- all physical properties of these states are identical under the appropriate exchange of labels 1,2,3.   In summary, as $\beta$ decreases through 4, there will be a discontinuous phase transition to a permutation symmetric point.   For $4-\beta \gtrsim 0.3$, we find that the only minimum of $E$ on the simplex is the permutation symmetric point, $q_i=1/3$.

In Section \ref{landausect} we will argue that the SSB transition for $n>2$ is generically discontinuous.

\subsubsection{More General Cases}
\begin{figure*}
\includegraphics[width=7in]{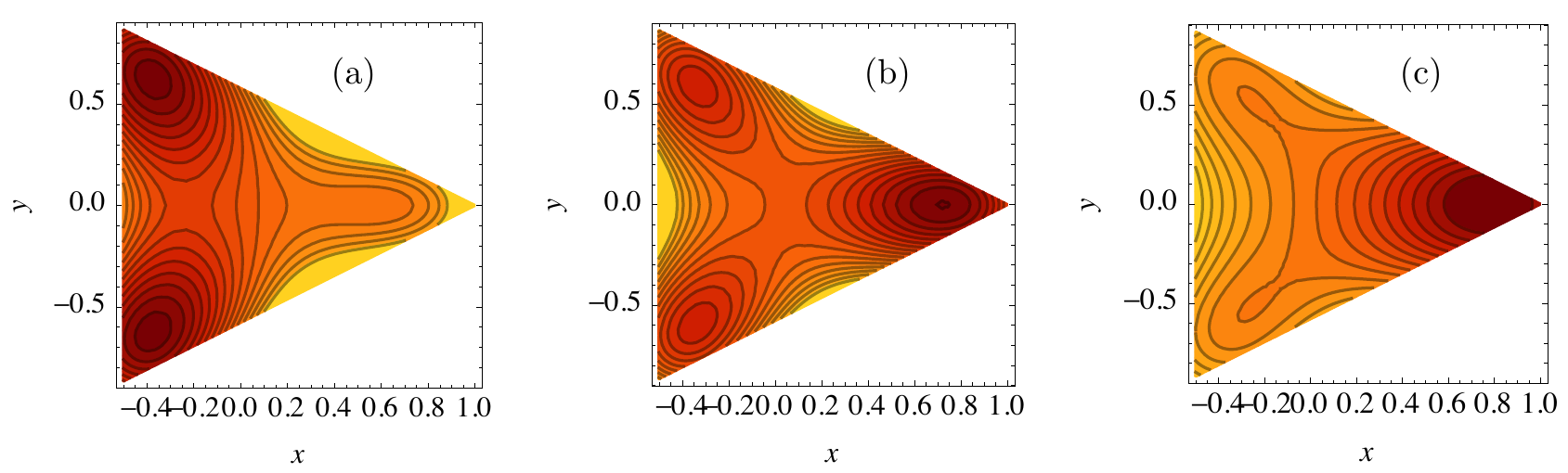}
\caption{(Color online.)  The energy landscape $E(x,y)$ when the average intrinsic utility of choice 1, in the SSB phase, differs by an amount $p$.    We use $\beta=4.1$ and (a), $p=0.03$, (b) $p=-0.01$ and (c)$p=-0.052$. Darker shading corresponds to $E$ smaller, and lighter shading to $E$ larger.  When there is a small positive 'price' $p$ attached to choice $1$ in (a), a plateau develops near the $q_1$ corner of the simplex, hence heavily favoring the other two choices. As the price of choice $1$ becomes more negative in (b) and (c), the minimum of $E$ tilts rapidly towards the choice $1$ corner.  Note that the requisite value of $p$ to completely shift the phase diagram is extremely tiny when compared to $\beta^{-1} \approx 0.25$ -- the typical spread in intrinsic utilities.}
\label{4bfig2}
\end{figure*}

Let us briefly discuss some cases where permutation symmetry is broken by the intrinisic utilities $F_i$'s.   One simple example of this is that we take $F_1$ and shift its argument by a value $p$, so that $F_1(u) = F(u+p)$.   This $p$ may correspond to some sort of ``price", as we will discuss in a later section.  In particular, as $p$ gets larger, then choice 1 becomes less and less attractive to the agents.    As a concrete example, we take $F(u)$ to be given by the logistic distribution.   What we find, as shown in Figure \ref{4bfig2}, is that the agents are rather sensitive to small changes in $p$.  

\begin{figure*}
\includegraphics[width=7in]{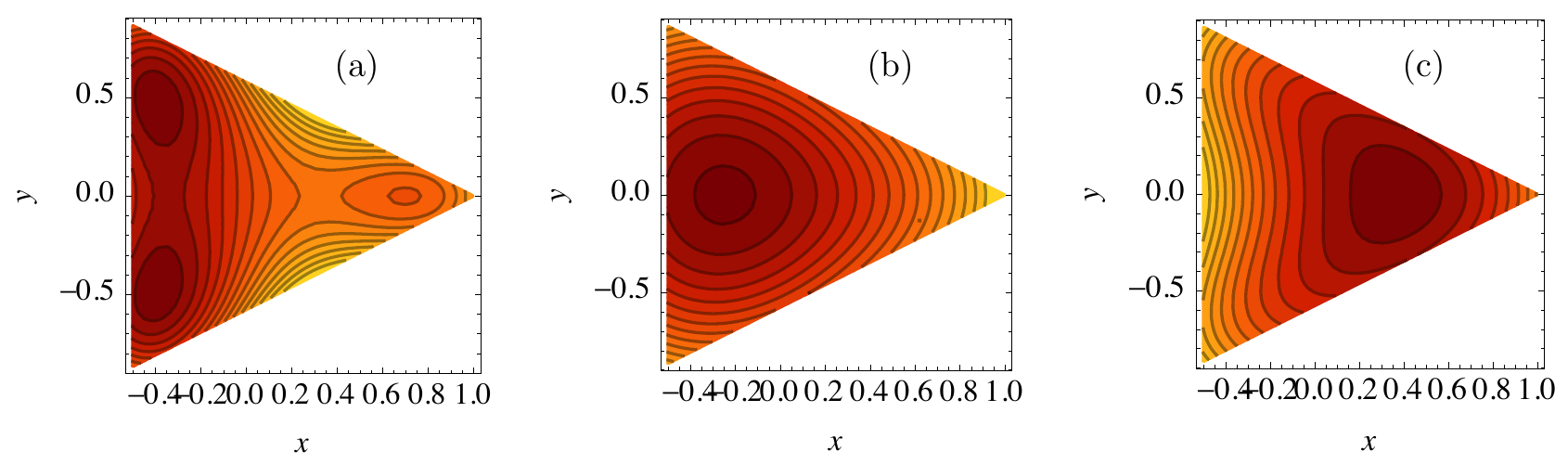}
\caption{(Color online.)  The energy landscape $E(x,y)$ for logistic decision making with permutation symmetry broken by distinct values of $\beta$ for different choices.  We take $\beta_2=\beta_3$ for simplicity.   (a) $\beta_1=6$, $\beta_2=3.5$, (b) $\beta_1=6$, $\beta_2=2$, and (c) $\beta_1=2$, $\beta_2=5$.  Darker shading corresponds to $E$ smaller, and lighter shading to $E$ larger.   In cases (a) and (b), choice 1 is much less heterogeneous, and we see that the minima shift towards favoring $q_2$ and $q_3$; the difference between (a) and (b) stems from the fact that the permutation symmetric fixed point for the binary decision problem between $q_2$ and $q_3$ is unstable in (a), and stable in (b).   Plot (c) shows a strong skew of the minimum towards favoring $q_1$ when choice 1 has more heterogeneity.}
\label{4bfig3}
\end{figure*}
Another example is to take all $F_i(u)$ given by the logistic distribution, but to take $\beta_1\ne \beta_2=\beta_3$.   In this case, we numerically find that the most stable fixed points correspond to the choices $i$ with the smallest value of $\beta_i$, as shown in Figure \ref{4bfig3}.   This can heuristically be explained as follows:  let us consider the limit where $\beta_1$ is finite, but $\beta_2=\cdots = \beta_n =\infty$ (the argument is general for any $n\ge 3$).   Then we can exactly compute \begin{equation}
G = -Q - \frac{\log \left(1+\mathrm{e}^{\beta(q_1-Q)}\right)}{\beta}
\end{equation}where $Q\equiv \max(q_2,\ldots, q_n)$.  Assuming permutation symmetry among $i=2,\ldots, n$, and using that (after taking $q$-derivatives of $G$) $(n-1)Q+q_1=1$, we conclude $q_1$ satisfies the equation \begin{equation}
q_1 = \frac{1}{1+\mathrm{e}^{\beta (1-nq_1)/(n-1)}}.
\end{equation}The left hand side of this equation is an increasing function of $q_1$, but the right hand side is decreasing; there is a unique solution.  It is easy to check -- e.g. by trying $q_1=1/n$ -- that the solution to this equation has $q_1>1/n$.   In this simple limit, we see that heterogeneity in choice 1 has broken permutation symmetry towards choice 1.   This heuristically explains how a choice with a wide variety of intrisinic utilities may end up obtaining a larger share of agents, even if the average utility gained from that choice is no better.   

A less rigorous, non-mathematical explanation of this is that for a choice with a wider variety of intrinsic utilities, most agents who are ``pinned" by their strong opinion to their favorite choice are in the state with more heterogeneity.   This pinning then encourages more of the remaining agents to also adopt this choice.

\section{Stability of a Fixed Point}\label{sec5}
We have seen repeatedly in our plots of the energy $E$ the disappearance and emergence of new local minima and maxima as paramters are tuned.   This is readily interpreted as the emergence of new \emph{phases}, exactly analogous to phases such as liquid water vs. solid ice.   

One of the most important questions is therefore -- if we are in a given phase, can this phase become unstable as we tune a given parameter?  If so, what is the endpoint of the instability -- where will the dynamics of opinion changes drive the system?   We will tackle these questions in this section -- first with macroscopic arguments based on a linear stability analysis of the energy $E$, and then justify these assumptions carefully by a microscopic analysis of the stochastic dynamics of individual agent state changes.

For the next two sections, we will assume that\footnote{This is equivalent to assuming that we are studying the ferromagnetic random field Potts model.} $\vec f = \vec q$.  This simplifies the presentation, although the logic carries through to the general case.

\subsection{Macroscopic Analysis}
%An important question of this paper concern the stability of various equilibria, which we now turn to in generality. 
First, we discuss the properties of fixed points at the macroscopic level of the effective energy.   Consider some fixed point $\vec q_*$.   We will subsequently show that this fixed point is only stable if it is a local minimum of the energy $E$.   For now, let us explore macroscopic consequences.  Taylor expanding the energy around $\vec q_*$: $\vec q = \vec q_* + \delta \vec q$:
\begin{equation}
E \approx E(\vec q_*) + \frac{1}{2}\sum_i \delta q_{i}^2 - \sum_{i,j} \frac{1}{2}\alpha_{ij} \delta q_{i} \delta q_{j} + \mathrm{O}(\delta q^3),
\end{equation}where we have defined
\begin{equation}
\alpha_{ij} \equiv -\frac{\partial^2 G}{\partial q_i \partial q_j}.
\end{equation}
Using the definition of $G$, we find that $\alpha_{ij}<0$ if $i\ne j$.\footnote{This follows from the positivity of $F_i$ and $F_i^\prime$.}
If we evaluate the matrix $\alpha_{ij}$ at an extremum of the energy, then the constraint that dynamics are constrained to the simplex implies that \begin{equation}
\sum_{i=1}^n \alpha_{ij} =0,  \label{sumalpha0}
\end{equation}which also gives us $\alpha_{ii}>0$. Note that $\alpha_{ij}$ is a symmetric matrix whose eigenvalues are therefore all real.

If we consider the dynamics of our system in real time, the simplest possible guess is that the dynamics are governed by relaxation to the ``ideal" value of $\vec q$, $-\partial G/\partial f_i$ (see Eq. (\ref{eqm1}));\footnote{Note that the appropriate units of time are undetermined.  We have chosen to scale the units of time so that the overall coefficient on the right hand side is 1.} denoting $\dot{q}_i \equiv \mathrm{d}q_i/\mathrm{d}t$,
\begin{equation}
\dot{q}_i = -\frac{\partial G}{\partial f_i} - q_i.  
\label{gradientG}
\end{equation}  
This differential equation is more carefully justified in Section \ref{realtimedynamics}. With $\vec f=\vec q$ here, we can also rewrite this as 
\begin{equation}
\dot{q}_i = -\frac{\partial E}{\partial q_i}.
\label{gradientflow} 
\end{equation}In this special case of interactions, we see that the dynamics can be written as a gradient flow, whereby the system relaxes to a local minimum of the energy.  This is analogous to Allen-Cahn relaxational dynamics \cite{allencahn}.

We will argue in the next subsection that the gradient flow behavior is indeed sensible from a microscopic perspective; in Appendix \ref{appgradflow}, we shall also show that gradient flow dynamics also holds for more general $\vec f(\vec q)$, but in a different ``coordinate system" $\vec q \rightarrow \vec \gamma(\vec q)$, with Eq. (\ref{gradientflow}) replaced by Eq. (\ref{gradientflowgen}).  

Assuming Eq. (\ref{gradientflow}), and $\vec f = \vec q$, if we linearize the energy near a fixed point, we find that
\begin{equation} 
\delta \dot{q}_i = \alpha_{ij}\delta q_j  - \delta q_i.  \label{deltaqdt}
\end{equation}
Evidently, if the eigenvalues of $\alpha_{ij}$ are all smaller than 1, the fixed point is stable, and if any eigenvalue is larger than 1, the fixed point is unstable.    Near a phase transition, when an eigenvalue of $\alpha_{ij}$ tends to 1, the dynamics will experience critical slowing down, as per usual.

We can also view the gradient flow in Eq. (\ref{gradientflow}) as analogous to the motion of a massless, positively charged particle in a viscous medium due to the ``electric potential" $E$. This electric potential can be visualized as arising from a background charge density $\rho$, obtained via Poisson's equation
\begin{align}
-\rho = \nabla^2 E &= n + \int \mathrm{d}u \sum_{i=1}^n F^{\prime\prime}_i(u-q_i)\prod_{j\neq i} F_j(u-q_j) \notag \\
&= n - \sum_i \alpha_{ii}
\label{poisson}
\end{align}
We see there is a uniform, constant contribution to $\rho$, arising from the contribution $\mathcal{H}+\vec q\cdot \vec f$, and a variable contribution to $G$.   In particular, it is readily seen from Eq. (\ref{poisson}) that the contribution to $\rho$ from $G$ can be interpreted as a weighted density of the likelihood that agents are about to switch their state.

Since $\alpha_{ii}\ge 0$ for any $i$, the contribution to the charge density from $G$ is always opposite to the constant negative background charge.   The competition between positive and negative charge densities therefore leads to the shape of $E$.  One implication of this is that as $\beta\rightarrow \infty$, $G(\vec f(\vec q))$ becomes sharper (Eq. (\ref{Gzerotemp})) and pushes the minima of $E$ away from the center of the simplex. In other words, a  more homogeneous intrinsic utility favors more strongly distorted outcomes, in agreement with the intuition that more homogeneous agents are more susceptible to social influence.

%\red{Reword this paragraph plz...}  Effectively, it corresponds to the divergence of neighboring MF market trajectories via $\rho=-\frac{1}{4\pi}\nabla\cdot \vec q$. %When $\rho=0$, $E$ satisfies the Laplace equation, just like the surface of a stretched membrane away from any support or rim. 
%In the common analogy with the stretched drumskin, the background charge $\rho$ behaves like an external repulsive mechanical influence. In this sense, it is a measure of how "relaxed" the MF market trajectory is. A region with high $\rho$ concentrates market trajectories, even if it does not trap any of them.  

\subsection{Avalanches and a Microscopic Perspective on Stability}
\label{avalanches}

In order to justify the assertions above, we now discuss the dynamics of avalanches.   By avalanche, we mean the following:  suppose that a single agent changes his state.   Subsequently, the values of $U_i + f_i$ change for the other nodes, and therefore other nodes may also change their state.   This leads to a cascade of state changes, which we call an avalanche.   We stress that the calculation below requires the assumption that, before the first agent changed his state, the system was at a fixed point.  This calculation is a generalization of similar results in the $n=2$ case in \cite{Lucas2013} -- see Appendix \ref{app:avalanches}, and is quite similar to  work done on the fiber bundle model \cite{pradhan}.

Let us compute the probability that any given agent switches from state $i$ to $j$, given that we alter the probability distribution from $q_i$ to $q_i+\delta_i$: \begin{widetext}
\begin{align}
\mathrm{P}(i\rightarrow j) &= \int \prod_{i=1}^n \mathrm{d}u_i F_i^\prime(u_i) \prod_{k\ne i} \Theta(u_i + q_i - u_k-q_k) \prod_{l\ne j} \Theta(u_j + q_j+ \delta_j - u_l - q_l - \delta_l) \notag \\
&= \int\limits^\infty_{-\infty} \mathrm{d}u_i F_i^\prime(u_i) \int\limits_{u_i + q_i - q_j - \max(0,\delta_j -\delta_i)}^{u_i + q_i - q_j} \mathrm{d}u_j  F_j^\prime(u_j) \prod_{k\ne i,j} F_k\left(\min(u_i+q_i, u_j+q_j+\delta_j-\delta_k)-q_k\right).   \label{pitoj0}
\end{align}
The first line in this expression is simply the statement that $U_i + q_i$ is maximal before the change in the probability distribution, and $U_j + q_j + \delta q_j$ is maximal after the change.   In the second line, we combined pairs of Heaviside $\Theta$ functions involving $u_k$ and integrated over $u_k$;  we also noted that the presence of a pair of $\Theta$ functions involving $u_i-u_j$ allows us to perform the $u_j$ integral as well.   Note that $\mathrm{P}(i\rightarrow j) >0$ if and only if $\delta_j > \delta_i$.  

Suppose further that only a finite number of agents change their state during the entire avalanche.  In this case, we can do a Taylor expansion of Eq. (\ref{pitoj0}).    In particular, if we are doing a Taylor expansion around $\vec \delta = \vec 0$, then the only possible contribution at first order comes from the Taylor expansion of the lower integrand on $u_j$ -- at leading order, the upper and lower bound are equal: \begin{equation}
\mathrm{P}(i\rightarrow j) \approx \int\limits_{-\infty}^\infty \mathrm{d}u_i F^\prime_i(u_i) (-\max(0,\delta_j - \delta_i)) \left[ F^\prime_j(u) \prod_{k\ne i,j} F_k(u)\right]_{u=u_i+q_i-q_j}.
\end{equation}
\end{widetext}
Using the definition of $\alpha_{ij}$ we conclude \begin{equation}
\mathrm{P}(i\rightarrow j) \approx (-\alpha_{ij}) \max(0,\delta_j-\delta_i),  \label{pitoj}
\end{equation}where we have used the definition of $\alpha_{ij}$ as a second derivative of $G$, and the fact that the integrand may be evaluated at $\vec \delta = \vec 0$, to obtain this answer.      Pleasingly, we see that $\alpha_{ij}$ admits a microscopic interpretation as the likelihood of state changes during an avalanche.

Given this linearized approximation near a fixed point, let us study the dynamics of avalanches.  This is similar to theory of multi-type branching processes \cite{Harris1963}, though with some important differences beyond first moments.    We work at a series of discrete time steps $t=0,1,2,\ldots$ and denote with $Z^t_i$ the change in the number of agents who are in state $i$, during time step $t$.  Note that \begin{equation}
\sum_{i=1}^n Z_i^t = 0,
\end{equation}and so $Z_i^t$ can be negative.   We can write \begin{equation}
Z_i^t = \sum_{j\ne i} \left(W_{ji}^t-W_{ij}^t\right)
\end{equation}where $W_{ij}^t$ is the number of nodes which will flip from $i$ to $j$ during time step $t$.   Based on the independence of the utilities of each node, we find that, defining \begin{equation}
Y^t_{ij} \equiv \max(0,Z^t_j - Z_i^t),
\end{equation}
\begin{widetext}
\begin{equation}
\mathrm{P}(W_{ij}^{t+1} = k) \approx \left(\begin{array}{c} N \\ k\end{array}\right) \left(|\alpha_{ij}| \frac{Y^t_{ij}}{N}\right)^k  \left(1-|\alpha_{ij}| \frac{Y^t_{ij}}{N}\right)^{N-k} \approx \frac{\left(|\alpha_{ij}|Y_{ij}^t\right)^k}{k!}  \mathrm{e}^{-|\alpha_{ij}|Y_{ij}^t}.  \label{pwij}
\end{equation}The distribution of $W_{ij}^{t+1}$ is thus Poisson with mean $|\alpha_{ij}|Y_{ij}^t$.

Suppose that $-Z^0_i = Z_j^0 = 1$ (i.e., the avalanche begins by a single node flipping from $i$ to $j$).    We obtain that\begin{equation}
\langle Z^{t+1}_i | t\rangle = \sum_{j\ne i} \langle W_{ji}^{t+1} - W_{ij}^{t+1} | t\rangle =- \sum_{j\ne i} \alpha_{ij}\left(Z_i^t - Z_j^t\right) = \sum_{j=1}^n \alpha_{ij} Z_j^t  \label{eqzt1}
\end{equation}
\end{widetext}
where $\langle \cdots | t\rangle$ means expectation values conditioned on the information of all state changes up to time $t$.   We only have to consider state changes that occured at time step $t$ to compute $Z^{t+1}_i$ because of the linearity of Eq. (\ref{pitoj}) -- only events that happen in the previous time step can lead to a change that would not have occurred previously.   In the last step, we employed Eq. (\ref{sumalpha0}).  We arrive at the nice result \begin{equation}
\langle Z^t_i \rangle =  \left(\alpha^t\right)_{ik}  Z^0_k \label{eq54}
\end{equation}where $\alpha^t$ is the $t^{\mathrm{th}}$ power of the matrix $\alpha$.    In retrospect, we do not really need to know the Poisson statistics of $W_{ij}^t$ to determine Eqs. (\ref{eqzt1}) and (\ref{eq54}).  They also follow from the fact that this stochastic process is Markovian (the statistics at time $t+1$ only depend on the value at time $t$).

If the largest eigenvalue of $\alpha$ is smaller than 1, the avalanche almost surely has finite size.   The total change $X_i$ in the number of agents in state $i$, given by \begin{equation}
X_i = \sum_{t=0}^\infty Z^t_i
\end{equation}has expected value \begin{equation}
\langle X_i\rangle = \left(1-\alpha\right)^{-1}_{ik} Z^0_k.  \label{eq26}
\end{equation}
This expression will be divergent as soon as the first eigenvalue of $1-\alpha$ tends to 0.   Such a divergence corresponds to the onset of instability.

\subsection{Real Time Dynamics}
%Let us conclude this section by arguing what the real time dynamics of this decision model would look like.   For simplicity, we assume that the system is near a stable fixed point.   To make the transition to continuous time easier, it is helpful to instead step forward time $t$ in steps of $\Delta \tau \le 1$, and to set the probability of changing state at each step (if a new state is more preferable) to be given by $\Delta \tau$.   On average, we thus expect the time we have to wait to change state to be $t=1$.   $\Delta \tau =1$ corresponds to the dynamics of the previous subsection.
%
%We can now go through and repeat the analysis of the previous subsection, replacing $\alpha_{ij}$ with $\alpha_{ij}\Delta \tau$ in Eq. (\ref{pitoj}) and subsequent formulae.   In the continuum limit $\Delta \tau \rightarrow 0$, we approximately only have one state change per time step.   this leads to the node-by-node differential equations
\label{realtimedynamics}
There is a natural alteration of the microscopic dynamics described above which possesses a simple continuum limit.  Let us consider a given agent $\alpha$.   Suppose that $\alpha$'s preferred state is $j$, but $x_\alpha(t)=i$.   Then we assume that in a discrete time step of size $\Delta \tau$, the probability of $\alpha$ transferring from $i$ to $j$ is given by $\Delta \tau$.   If $\Delta \tau = 1$, then we recover the dynamical rules of the previous subsection.    If $\Delta \tau < 1$, then \begin{subequations}\begin{align}
\mathrm{P}(x_\alpha(t+\Delta\tau)=j) &= \left\lbrace\begin{array}{ll} 1&\  x_\alpha(t) = j \\ \Delta \tau &\ x_\alpha(t) \ne j \end{array}\right., \\
\mathrm{P}(x_\alpha(t+\Delta\tau)=i\ne j) &= \left\lbrace\begin{array}{ll} 1 - \Delta \tau &\  x_\alpha(t) = i \\ 0 &\ x_\alpha(t) \ne i \end{array}\right..
\end{align}\end{subequations}
In the limit $\Delta\tau\rightarrow 0$, we may treat $t$ as a continuous variable, and these update rules reduces to the differential equation
\begin{equation}
\frac{\mathrm{d}}{\mathrm{d}t}\mathrm{P}(x_\alpha=i) = \delta_{ij}-\mathrm{P}(x_\alpha=i).  \label{eq57}
\end{equation} Averaging over all nodes $\alpha$, Eq. (\ref{eq57}) becomes \begin{equation}
\dot{q}_i = \mathrm{P}(i\text{ optimal}) - q_i
\end{equation}which is simply Eq. (\ref{gradientG}).   
%Formally speaking, the avalanche analysis was only carried out in the linear response regime, but it is reasonable that this equation holds outside this regime as well, for reasonable microscopic dynamics.

\section{Permutation Symmetric Models }
\label{SSBpermsymm}
In this section, we shall perform a more detailed study on permutation symmetric models, which correspond to \begin{equation}
F_i(u) \equiv F(u)
\end{equation}for each $i$.    These are the simplest models to analyze, and they turn out to be interesting in their own right.  It is  straightforward to see that the permutation group $\mathrm{S}_n$, whose action interchanges the labels on $q_i$, is a symmetry of the energy $E$.    The most obvious question then becomes:  under what conditions (and to what subgroups) does the permutation symmetry spontaneously break.  Correspondingly, when will interacting agents spontaneously begin to prefer certain choices over others, despite the fact that they are all inherently the same?\footnote{\cite{Salganik2006} referred to this phenomenon as ``unpredictability".   Although the state which the system spontaneously picks cannot be predicted, ``physical properties" of the resulting state can be.   As an analogy, the Ising model on the square lattice undergoes a phase transition at low temperatures:  the magnetization is spontaneously either up or down.   Nonetheless, the physical properties of the magnet are identical for each phase.   Decision making where the phase diagram is unpredictable is described in \cite{Lucas2013b}.}   
This will be the main theme of this section.    Although a real world market may not have permutation symmetry, these models serve as solvable toy models where we may disentangle the effects of intrinsic heterogeneity between choices, and the effects of social interactions.   Given that experiments \cite{Salganik2006} suggest the latter is relevant in real world decision making, this is a natural question to study.   We emphasize the importance of SSB for non-specialists in Appendix \ref{ssbintro}.

\subsection{Stability of the Permutation Symmetric Fixed Point}
Directly analyzing the stability of the permutation symmetric fixed point is straightforward, and so we begin here.  By permutation symmetry, $\alpha_{ij}$ must be of the form \begin{equation}
\alpha_{ij} = a+b\delta_{ij}
\end{equation}for constants $a$ and $b$.   Using Eq. (\ref{sumalpha0}) we find that \begin{equation}
an+b=0. \label{defb}
\end{equation}We also know that the dynamics is constrained to the simplex, which means that we only need consider eigenvectors $\delta \vec q$ for which $\delta q_1 +\cdots + \delta q_n =0$.   Since the $n-1$ eigenvectors which satisfy this constraint have eigenvalue $b$, we conclude that the stability of the fixed point is controlled entirely by this one eigenvalue.  We can compute $b$ by computing a diagonal element of
 $\alpha$: \begin{equation}
\alpha_{11} = a+b = \frac{n-1}{n}b.   \label{alpha11}
\end{equation}If $b<1$, the permutation symmetric point is stable; if $b>1$, it is unstable, and if $b=1$, higher order corrections are required.

One of the most important questions to ask is what effect the addition of more choices has on stability:   do more choices make the permutation symmetric point more or less stable?   It turns out that the answer depends on the specific choice of $F(u)$, and on any possible scaling limits of $F(u)$ as $n$ gets large.   We briefly discuss this question, as well as aspects of the large $n$ limit in Appendix \ref{largenapp}.   Whether or not a utility based model of decision making has any real-world relevance in the large $n$ limit is a question to take seriously, however.

\subsection{Landau Theory}
\label{landausect}
Given an understanding of the stability of a permutation symmetric point, let us now return to a statement we claimed earlier -- it is non-generic for a permutation symmetry breaking phase transition to be continuous.   We now sketch out why this is so, leaving details to Appendix \ref{landauapp}.

Near a permutation symmetric point, if we had a continuous phase transition, then we can expand out the energy $E$ as a Taylor series in $\delta q_i = q_i - 1/n$: \begin{align}
E &\approx -\zeta \sum_{i=1}^n \delta q_i^2 - \xi \sum_{i=1}^n \delta q_i^3 + \omega \sum_{i=1}^n \delta q_i^4 + \psi \left(\sum_{i=1}^n \delta q_i^2\right)^2 \notag \\
&\;\;\;  + \mu \sum_{i=1}^n \delta q_i  \label{landaueq}
\end{align}As we explain in the appendix, this is the most general form of $E(\delta \vec q)$ consistent with permutation symmetry.  The final $\mu$ term serves as a Lagrange multiplier, enforcing that we are on the simplex.  When $\zeta < 0$,  the permutation symmetric phase is stable;  when $\zeta>0$, it is unstable.  For the $\vec f = \vec q$ model, we prove $\xi>0$ in Appendix \ref{landauapp}.  A thorough analysis there reveals that for $|\xi|>0$, and $\zeta\rightarrow 0$, the minima of $E$ occur when $\delta q_1 = \cdots \delta q_p \ne \delta q_{p+1} = \cdots = \delta q_n$ (up to the action of $\mathrm{S}_n$), and that $\delta q_1$ is finite even as $\zeta \rightarrow 0$.   This demonstrates that within Landau theory, the phase transition is discontinuous so long as $\xi \ne 0$ (setting $\xi=0$ requires fine tuning one parameter).  This result is not particularly strange; it is well known that the ferromagnetic (non-random field) Potts model \cite{kihara, wu} has a discontinuous phase transition for $n>2$ at mean field level; here we see it holds at zero temperature with random fields, and with a very general class of interactions.   Thus, we see from simple symmetry-based arguments that nearly every symmetry-breaking phase transition in the $n>2$ models should be discontinuous.   This has important consequences for economics -- for example, without a large amount of heterogeneity in intrinsic utilities (relative to social utilties), nearly every phase transition will be discontinuous, suggesting the prevalence of market crashes.

\subsection{Unimodal Distributions}
It is of interest to us to analyze the symmetry of the global minima of $E$ when the permutation symmetric fixed point is unstable.   Consider the simple case where $F(u)$ describes a ``unimodal" distribution which is (roughly speaking) clustered around a single point.   Prototypical examples are the uniform distribution Eq. (\ref{uniformdist}), the logistic distribution Eq. (\ref{fdapp}), or a Gaussian distribution $F(u) = (1+ \mathrm{erf}(u/\sigma))/2$.   

We have already described in detail the energy landscape of the logistic ternary decision model in a previous section.   In this model, we found that the permutation symmetry group was broken from $\mathrm{S}_3$ to $\mathrm{S}_2$  -- i.e., we picked a preferred choice (e.g. $q_1^* > q_2^* = q_3^*$).   

It is qualitatively easy to understand why this should be the case with the following simple argument: suppose that there is no heterogeneity in the quenched disorder (intrinsic utility):  \begin{equation}
F(u) = \Theta(u) = \left\lbrace \begin{array}{ll} 1 &\ u\ge 0 \\ 0 &\ u<0 \end{array}\right..
\end{equation}This corresponds to the limit $\beta \rightarrow \infty$, in our logistic model.  Then, using $\delta q_i \equiv q_i - 1/n$, let us evaluate the energy $E$ constrained to the space \begin{equation}
\sum_{i=1}^n \delta q_i^2 = \text{constant}.
\end{equation}Remember that the constraint $\delta q_1+\cdots+\delta q_n=0$ still applies.    The quadratic term in $E$ is clearly a constant on the sphere, and so the only quantity of interest is $G$, which can be easily evaluated:\begin{equation}
G(\delta q_1,\ldots,\delta q_n) = -\max(\delta q_1,\ldots,\delta q_n).
\end{equation}We see therefore that for any fixed distance from the permutation symmetric fixed point, the minima of $G$ (and thus of $E$) are at points where (without loss of generality)\begin{equation}
\epsilon_1 = -(n-1)\epsilon_i, \;(i=2,\ldots,n).
\end{equation}This corresponds to spontaneous symmetry breaking to the subgroup $\mathrm{S}_{n-1}\subset \mathrm{S}_n$, where exactly one choice becomes preferential over the others.    In this deterministic model, one can in fact check that the minima of $E$ occur precisely when $q_1=1$.

The role of disorder in $F(u)$ is to push (in many cases) $q_1$ to a slightly smaller value.   Of course, in some cases, we have seen that disorder is strong enough that $q_1$ is pushed all the way to the permutation symmetric point:  $q_1=1/n$.  However, if all choices are drawn from unimodal distributions, then our numerical analyses suggest though no further symmetry breakings are possible -- only one choice becomes more favored over the others.    For a rigorous analysis in the large $\beta$ limit of logistic models, see Appendix \ref{appendixg}.

\subsection{Bimodal Distributions}
 A more interesting case to consider is a bimodal distribution, where there are two sharp peaks in $F(u)$.    Curious phase diagrams are known to arise in ``$\mathrm{O}(n)$" spin models (which do not have discrete choices) subject to bimodal random fields in certain directions \cite{galam82}.  We will see that rich behavior can arise here as well.

The simplest example of this is to take \begin{equation}
F(u) = p\Theta(u) + (1-p)\Theta(u-u_0).
\end{equation}We can think about this intuitively as follows: for each choice, one dislikes it with probability $p$, and likes it with probability $1-p$.

Without loss of generality we set $q_1 > q_2 > \cdots \ge  q_n$.   Then we can compute $G$ very straightforwardly: \begin{equation}
G = \int\limits_{-R}^R \mathrm{d}u \prod F(u-q_i) = \int\limits_{q_1}^R \mathrm{d}u \prod F(u-q_i).
\end{equation}The coefficient $R$ here is a regulator, and will not affect the answer.   Let us suppose that $q_{m+1}=0$, but $q_1>q_2>\cdots > q_m >0$.   Note that this requires $u_0<q_1\le 1$.   Then the integral is easy to compute: \begin{widetext} \begin{equation}
G = p^m \left(q_m + u_0 - q_1\right) + p^{m-1}\left(q_{m-1}-q_m\right) + \cdots + p\left(q_1-q_2\right) + R-q_1.
\end{equation}
\end{widetext}
From this equation and Eq. (\ref{eqm1}) we can straightforwardly deduce that: \begin{subequations}\label{bimodalqs}\begin{align}
q_1 &= 1-p + p^m, \\
q_i &= p^{i-1}(1-p)  \;\;\;\; \text{ for }1<i\le m.
\end{align}\end{subequations} Note for this solution to be allowed we require that $q_m+u_0 \ge q_1$ or\begin{equation}
p^m \ge  u_0 - 1+p \ge  2p^m - p^{m-1}  \label{hierconstraint}
\end{equation}

There is a very simple way of understanding Eq. (\ref{bimodalqs}).    If we like the most popular choice (probability $1-p$), we will certainly go with that; if we don't, we ask if we prefer the next one (probability $p(1-p)$), etc.   Finally, if we dislike all of the $m$ choices which are represented at mean field level, we simply go with the most popular choice, as we dislike them all. 

Let us suppose that this condition is obeyed.   Then, using that, within a local patch, anywhere on the simplex, the energy function $E$ is a quadratic polynomial of the $q_i$s,  the value of the total energy $E$ on a solution with symmetry breaking as above is \begin{widetext} \begin{align}
E_m &= p^m u_0 - \frac{1}{2}\left[\left(1-p+p^m\right)^2 + (1-p)^2\sum_{k=1}^{m-1} p^{2k}\right]= p^m u_0 - \frac{1}{2}\left[\left(1-p+p^m\right)^2 +\frac{p^2\left(1-p^{2m-2}\right)(1-p)}{1+p}\right].
\end{align}We then note that\begin{align}
E_{m-1} - E_m = (u_0 - 1 +p ) p^{m-1}(1-p) -  p^{2m-1}(1-p) = p^{m-1}(1-p)(u_0-q_{1,m}) < 0 .
\end{align}
\end{widetext}
We conclude that $E_{m-1}>E_m$, and thus \emph{higher levels of spontaneous symmetry breaking is always ``more stable"}.   We have to be slightly careful about discussing global stability based solely on energetic considerations, as generically in a social model there is no reason that dynamics have to favor lower energy minima over others \cite{Lucas2013b}, but this is certainly suggestive that strongly permutation symmetry broken states are the endpoint of dynamics.   A more thorough analysis would compute the $\alpha_{ij}$ matrix at each permutation symmetry broken point, but this is cumbersome and we will not do it here.

\begin{figure*}
\includegraphics[width=7in]{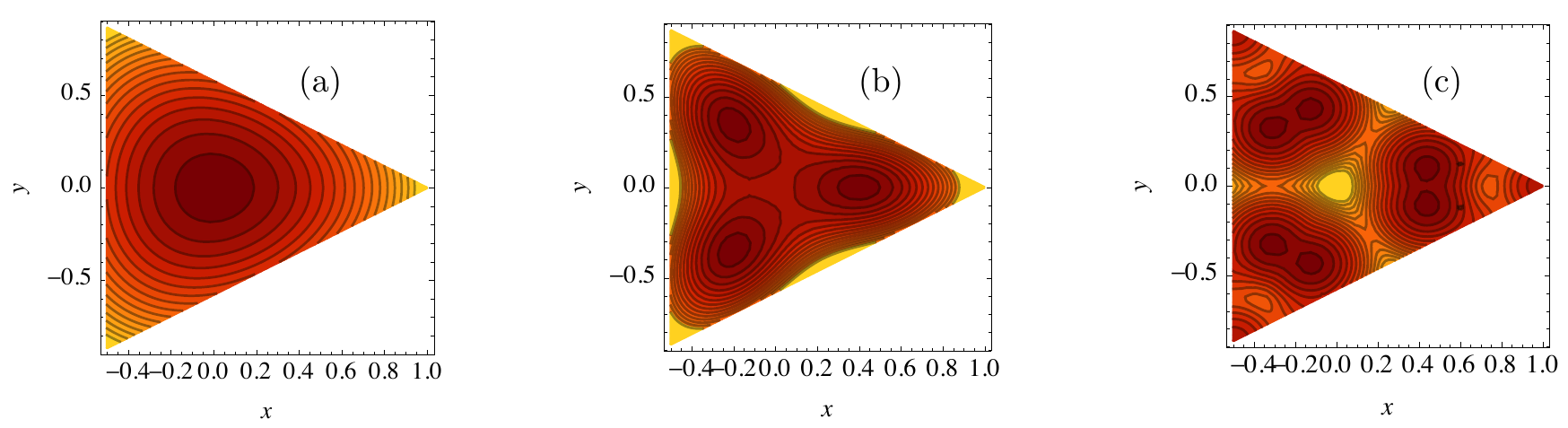}
\caption{(Color online.)  An example of SSB and hierarchy formation using $F(u)$ given by Eq. (\ref{bimodalfigeq}), with $p=0.5$, $u_0=0.73$, and (a) $\beta=3$ and $\mathrm{S}_3$ symmetry at the global minimum, (b) $\beta=8$, and $\mathrm{S}_2$ symmetry at the global minima (c) $\beta=50$ with no symmetry at the global minima.  Decreasing heterogeneity leads to further breaking of permutation symmetry.  We show contour plots of $E(x,y)$, with $(x,y)$ a barycentric representation of the simplex; darker shades correspond to smaller values of $E$, and lighter shades to larger values of $E$.}
\label{bimodalfig}
\end{figure*}

Finally, we note that Eq. (\ref{hierconstraint}) requires increasingly fine-tuned $u_0$ and $p$ to break to smaller subgroups of $\mathrm{S}_n$.   However, the formation of a hierarchy with $\mathrm{S}_n$ broken at least twice does not require particularly strong tuning.     One can also smooth out the bimodal distribution, e.g. by replacing step functions with logistic functions: \begin{equation}
F(u) = \frac{p}{1+\mathrm{e}^{-\beta(u-u_0)}} +  \frac{1-p}{1+\mathrm{e}^{-\beta u}}.  \label{bimodalfigeq}
\end{equation}Repeated SSB and the formation of hierarchies is still possible, as we show in Figure \ref{bimodalfig}.  In fact, by increasing the parameter $\beta$, we find a series of transitions at which the permutation symmetry breaks further and further (at the global minimum of $E$).  We expect that a similar phenomenon of repeated symmetry breaking as $\beta$ increases will happen for $n>3$.

\subsection{An Opt-Out Option}
In this section, we will suggest a way of using our decision model to model the behavior of markets where a 'no buy' option is allowed. Since we still have a $\mathrm{S}_{n-1}$ symmetry, we can do this by letting choices $1,\ldots, n-1$ refer to undifferentiated sellers which an individual is buying a product from.  Alternatively, these choices could be $n-1$ identical (on average) products, but which are distinguishable to individual consumers. we will let the final 'choice' $n$ correspond to the possibility that a buyer has opted out.  

We assume that the ``no buy" choice has a fixed utility:\begin{equation}
F_n(u) = \Theta(u)
\end{equation}and \begin{equation}
f_n = 0.
\end{equation}For notational ease, let $m=n-1$.   Note that we have used our freedom to shift the overall additive constant in social utility $f_i$ to set the utility of not buying to be 0.

Of course, we will want to include some aspect of ``pricing" in our model.    We will achieve this by heuristically defining $\vec f$ as follows (for the remainder of this section, the vector index will only be over indices 1 to $m-1$): \begin{equation}
\vec f=  \vec q - \vec p,
\label{pricevector}
\end{equation}where $\vec p$ is a ``price" vector.  Formally at this point it only corresponds to some external ``driving" of the system, which we will associate with changes in the prices of various sellers.   

 The energy to be minimized is \begin{equation}
E = \frac{1}{2} \sum_{i=1}^m q_i^2  + G(\vec q - \vec p)
\end{equation}where the form of the $G$ function is \begin{align}
G &= \int\limits \mathrm{d}u   \Theta(u)  \prod_{i=1}^{m} F(u-q_i+p_i) \notag \\
&= \int\limits_0^\infty \mathrm{d}u  \prod_{i=1}^{m} F(u-q_i+p_i)
\end{align}where again, a simple regularization of $G$ is required.   Note that $G$ is independent of $q_n$.    We will define $\alpha_{ij}$ similarly as before, but we will usually neglect $\alpha_{in}$ (this drops the constraint Eq. (\ref{sumalpha0}), when the sum is restricted to $i=1,\ldots,m$). 

As a simple example, we look at permutation symmetric fixed points with $\mathrm{S}_m$ symmetry, so that all $q_i=q$ and all $p_i=p$ for $i=1,\ldots,m$.   This corresponds to solutions which satisfy \begin{equation}
q = \int\limits_0^\infty \mathrm{d}u  \frac{1}{m}\frac{\mathrm{d}}{\mathrm{d}u}\left[F(u-q+p)\right]^m = \frac{1-F(p-q)^m}{m}.
\end{equation}
Symmetry strongly constrains the form of $\alpha$ at such a fixed point:  just as before, we have $\alpha_{ij} = a+b\delta_{ij}$, but now neglecting the $n$ index there is no constraint relating $a$ to $b$.   There are two eigenvalues of this matrix on the simplex now:   the eigenvalue $a$ corresponds to the scenario where buyers simply shuffle between choices, but none enter or leave the market, and corresponds to perturbations with $\epsilon_1+\cdots +\epsilon_m = 0$.    The eigenvalue $b+na$ corresponds to retaining permutation symmetry ($\epsilon_1=\cdots = \epsilon_m$) but having agents leave or enter the marketplace.   Since \begin{equation}
a = -\int\limits_0^\infty \mathrm{d}u F^{\prime}(u-q+p)^2 F\left(u-q+p\right)^{m-2} < 0
\end{equation}we conclude that \emph{spontaneous symmetry breaking is always the dominant instability} of a permutation symmetric fixed point.    

\begin{figure*}
\includegraphics[width=7in]{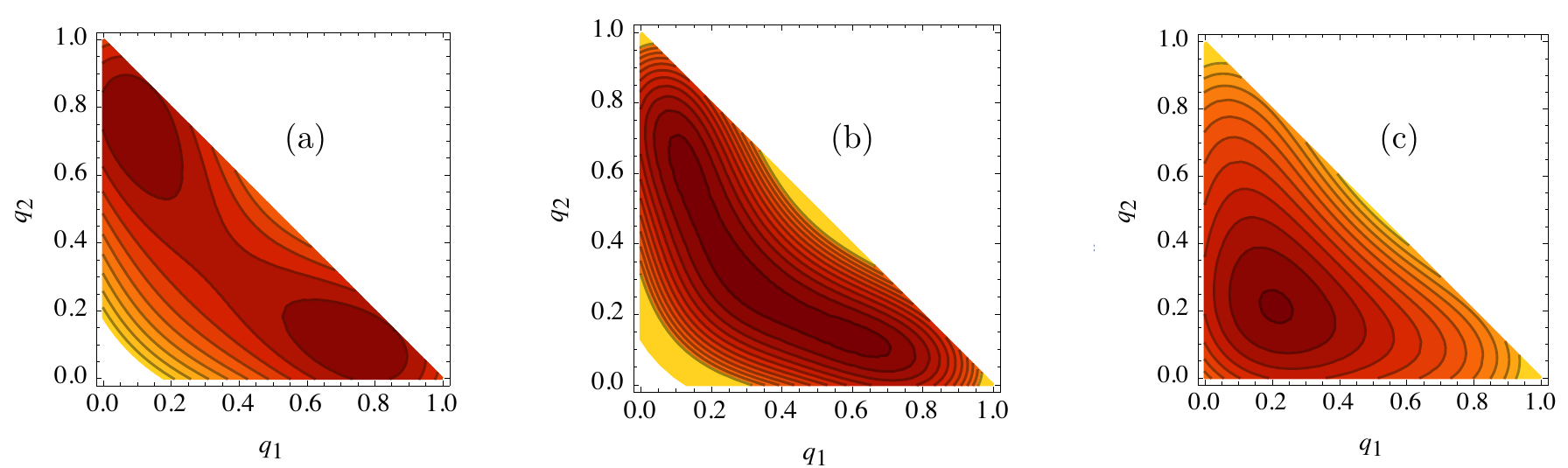}
\caption{(Color online.)  An example of ``opt out" multiple choice decision making with $m=2$ choices of sellers/products.   Note that the upper right corner has $q_1+q_2>1$, which is forbidden by the simplex constraint.   We used a logistic distribution for $F$ with $\beta=4$. Displayed are contour plots of $E(q_1,q_2)$ with prices $p_1=p_2=p$ (defined in Eq. \ref{pricevector}) for (a) $p=0.4$, (b) $p=0.44$, (c) $p=0.5$;  darker shades correspond to smaller values of $E$, and lighter shades to larger values of $E$.  As $p$ increases, we can see the phase transition from a SSB phase to a permutation symmetric phase.  This can be understood heuristically: at large $p$, only the agents with strong preferences for products are in the market, and there are not enough remaining agents to have comparably large social utilities.}
\label{5dfig}
\end{figure*}

If we use the logistic distribution for $F(u)$, equilibria are found by solving \begin{equation}
q = \frac{1}{m}\left[1-  \frac{1}{\left(1+\mathrm{e}^{-\beta(p-q)}\right)^m}\right].
\end{equation}
Bistability is only possible if the $q$-derivative on the right hand side takes values larger than 1, which occurs if there is a $q$ such that \begin{equation}
1<F^\prime(p-q) F(p-q)^{m-1}.
\end{equation}
We show an explicit example with $m=2$ and SSB in Figure \ref{5dfig}.

Let us briefly discuss an empirical application of the above observation.   Suppose one has a market with distinct products (or sellers), but with strong regulation, so that the price of the market is fixed.    This model predicts that \emph{if} the permutation symmetric point ever becomes unstable, as the regulator lowers the price in the market, interactions will drive the system to a symmetry broken point.   Although in principle this should be readily observable, in practice the permutation symmetric assumption is likely too strong.    It may be the case, however, that this heuristic observation of regulated markets being more unstable to herding than symmetry-preserving crashes, may be observable given aggregated economic data.

\section{Complexity on Graphs}
\label{sec:complexity}
In this section, we discuss the emergence of \emph{complexity} -- an exponential number of solutions to the equations of state -- when this decision model describes agents interacting via  a graph.   Statistical physics on random graphs has played an important role in the development of interdisciplinary physics, due to (reasonable) hope that the physics of models on random graphs captures key insight into the behavior of realistic social systems on realistic social networks \cite{Newman2003, Barrat2008}.   \cite{Lucas2013, Lucas2013b} discussed aspects of complexity for binary decision models on graphs, so we conclude with a brief, analogous discussion for $n>2$.

It is straightforward to extend our model to allow for decision making on graphs.  Let us consider a graph $G=(V,E)$, with $V$ the vertex set (we label vertices $u,v,\ldots$) and $E$ the edge set, consisting of undirected, unweighted edges between two nodes.   The edge between nodes $u$ and $v$ is denoted with $(uv)$.  The degree of node $v$, or the number of edges connecting to node $v$, is a positive integer $k_v$;   we denote with $\langle k\rangle $ the average number of edges per node.    As usual, we will consider a ``random graph" limit where the number of nodes tends to $\infty$, but the degree distribution (and thus $\langle k\rangle$) is fixed, and we assume that there are no correlations in the likelihoods of edges between nodes of different degree.  Denote with $N$ the total number of nodes.

   The only change required to our model is as follows:   for each node $v$, we define $\vec q_v$ to be the probability that a neighbor of node $v$ is in a given state: \begin{equation}
q_{v,i} \equiv \mathrm{P}\left(x_{u,i} = 1\; |\; (uv)\in E\right).
\end{equation}We then replace Eq. (\ref{eq2}) with \begin{equation}
V_{v,i} = U_{v,i} + f_i(\vec q_v).
\end{equation}The mean-field limit of these equations reduces to the formalism described above, and corresponds to the limit where $k_v\rightarrow \infty$.   We assume that we are expanding around a stable fixed point of the MF equations.  For the remainder of ths section, we will work in the limit where $n/k_v$ is small.   This allows us to reliably do perturbation theory around a MF state, where node $v$ sees $\approx k_v q_i \gtrsim 1$ of its neighbors in state $i$.\footnote{If this number can be small compared to 1, then random fluctuations in the states of neighbors become important -- see \cite{Lucas2013}.}

\begin{figure}
\includegraphics[width=2.7in]{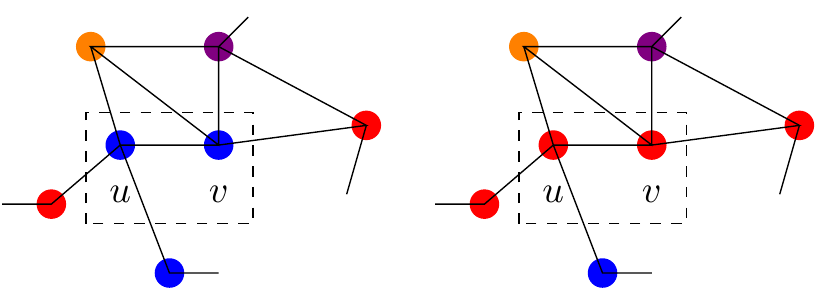}
\caption{(Color online.)  A connected pair of nodes $u$ and $v$, for which there is one solution to the equations of state where they are both in state $i$, and a second solution where they are both in state $j$.   Colored circles denote nodes in the graph; different colors are different states.   All other nodes have identical states in each of these two solutions.   In the thermodynamic limit, the presence of these clusters leads to an exponential number of solutions to the equations.}
\label{thoughtfig}
\end{figure}

It is simplest to understand the emergence of complexity with a simple ``thought experiment" \cite{Lucas2013}.   Let us ask for the probability that there is a pair of nodes $u$ and $v$, which are connected, and for which there are two solutions to the equations of state:  one where $u$ and $v$ are both in state $i$ (and $w\ne u,v$ are in some irrelevant states $x_w$), and one where $u$ and $v$ are both in state $j$, but for $w\ne u,v$, the states are unchanged:  see Figure \ref{thoughtfig}.  Using Eq. (\ref{pitoj}), so long as $k_{u,v}\gg 1$, we find that the probability for this occuring is $4|\alpha_{ij}|^2/k_uk_v$.  When $n=2$, that is the end of the story, but for $n>2$ it is a bit more subtle.    There is also a possibility that node $u$ or $v$ starts in state $k$ intsead of $i$, or that node $u$ or $v$ starts in state $i$ but ends in state $l$.   Accounting for these two possibilities as well, one finds the total probability that there are two solutions to the equations of state where nodes $u$ and $v$ take on different values in each solution, but all other nodes take on the same values in both solutions, is\footnote{The second line follows from the first, using the identities that the sum of the entries of any row of $\alpha_{ij}$ vanishes.} \begin{align}
\mathrm{P}(uv) &= \frac{4|\alpha_{ij}|^2}{k_uk_v} + \sum_{l \ne i,j} \frac{2|\alpha_{ij}|(|\alpha_{lj}|+|\alpha_{il}|)}{k_uk_v} \notag \\
&= \frac{2|\alpha_{ij}|(|\alpha_{ii}|+|\alpha_{jj}|)}{k_uk_v}
\end{align}Next, let us pick a node $v$, chosen uniformly at random in the graph.   What is the probability that there exists one solution to the equations of state where $z_{v,i}=1$, and another solution where $z_{v,j}=1$?   The answer is given by \begin{equation}
 \mathrm{P}(i,j) = \frac{2|\alpha_{ij}|(\alpha_{ii}+\alpha_{jj})}{\langle k\rangle}.  \label{eq69}
\end{equation}
In the macroscopic limit such a change in $\vec q$ is not noticable for a single pair of nodes, but because there is a finite probability for any given pair of nodes that multiple solutions exist, there are an exponential number of possible solutions to the equations of state, and there is a macroscopic spectrum of allowed values for $\vec q$.    

Note that when $\langle k\rangle \rightarrow \infty$, the complexity phenomenon is not present.  This has to happen, because there is no complexity on a complete graph.

There is a simple mathematical framework called the Thouless-Anderson-Palmer (TAP) equation \cite{tapcite, DeDominicis2006} (related to a mathematical technique called belief propagation \cite{mezard}) which allows us to make the simple calculation above more formal and to include the possibility of clusters with more than 2 nodes.   In particular, we can consider the possibility of clusters of arbitrary size.  These equations can, in principle, be exactly treated non-perturbatively when the graph is approximated to be a tree, as was done in the simple case $n=2$ earlier \cite{Lucas2013}.  Exact solvability is related to a nested structure of probability measures, which in turn follows from the fact that there is a unique path between any two nodes on a tree.

Our strategy for computing the generalization of Eq. (\ref{eq69}), accounting for the possibility of arbitrary sized clusters on a locally tree-like graph, is as follows.   Let us define with $\xi_{ij}$ the probability, accounting for the possible behavior of other nodes, that a given node $v$'s neighbors will flip from $i$ to $j$, given that either one of its neighbors flipped from $i$ to any other state, or from any other state to $j$.   (In particular, $\xi_{ij} \ne \alpha_{ij}$, because it may be possible that a node will only flip when two of its neighbors have flipped.)  Then, by definition: \begin{align}
\mathrm{P}(i,j) &=\frac{2|\alpha_{ij}|}{\langle k\rangle} \left\langle \delta q_{v,j} - \delta q_{v,i} \right\rangle \notag \\
&=\frac{2|\alpha_{ij}|}{\langle k\rangle}\left(\sum_{l\ne i}\xi_{li}+\sum_{l\ne j}\xi_{lj}\right).  \label{eq70}
\end{align}This is our effective belief propagation equation or TAP equation.  The factor of 2 in front accounts for the fact that we can either start from $i$ and end in state $j$, or vice versa, just as before.   $|\alpha_{ij}|/\langle k\rangle$ is the coefficient of proportionality in the probability that node $v$ would flip from $i$ to $j$, given the value of $\delta q_{v,j}-\delta q_{v,i}$, calculated under the assumption that node $v$ flipped from $i$ to $j$.   By definition of $\xi_{lm}$, the expectation value of $\delta q_{v,j} - \delta q_{v,i}$ can be easily written down.   
%Because the values of $U_{u,i}$ are drawn i.i.d. for each node $u$, we conclude that the probability measure on $U_{v,i}$ is independent of the probability measure on the avalanches, thus allowing us to multiply averages in computing $\mathrm{P}(i,j)$.   

So we have to simply find an expression for $\xi_{ij}$.\footnote{We cannot directly use Eq. (\ref{eq26}), because we must be careful to count the number of nodes correctly.   The subtlety is as follows:  in the computation of $Z^i_t$ before, we noted that it was twice as likely as $|\alpha_{ij}|$ for a node to flip aligned with its neighbor $i\rightarrow j$.}  To do this, we use the following recursive equation: \begin{equation}
\xi_{ij} = |\alpha_{ij}|\left(1+\sum_{l\ne i} \xi_{li} + \sum_{l\ne j} \xi_{lj}\right),
\end{equation}The factor of $|\alpha_{ij}|$ comes as usual from Eq. (\ref{pitoj}) -- the factors of $1+\sum \xi$ come from both the neighbor we engineered to flip, and then the response of all other neighbors.\footnote{Recall the following minor subtlety -- this equation relies on the factorization of the probability measure.   This does not happen if there are any cycles in the graph, because two neighbors of a node $v$ may feel each other's influence even if $v$ does not change.   The assumption that our graph is locally tree-like allows for factorization of the measure, and thus makes this equation exact.}    It is now straightforward to solve for $\xi_{ij}$.  Denote the matrix \begin{equation}
\mathcal{M}_{ij,kl} = \left\lbrace\begin{array}{ll} (|\alpha|_{ij})^{-1} -2 &\ ij = kl \\ -1 &\ ij,kl\text{ have 1 letter in common} \\ 0 &\ \text{otherwise} \end{array}\right.
\end{equation}where we consider $ij,ji$ to be the same index.   Then it is easy to see that we simply have to solve the linear algebra problem \begin{equation}
\sum_{kl} \mathcal{M}_{ij,kl}\xi_{kl} = 1 \label{eq80}
\end{equation}to determine $\xi_{ij}$, and thus determine $\mathrm{P}(i,j)$.  It is not guaranteed that this equation has a physical solution -- this absence may correspond, e.g., to the fact that the original fixed point was not stable (and thus clusters trivially percolate through a finite fraction of the entire graph).  In general the resulting expression will be quite complicated; in the limit where $|\alpha_{ij}| \ll 1/n$,  we can approximate that $\xi_{ij} \approx |\alpha_{ij}|$, and $\mathrm{P}(i,j)$ is simply given by Eq. (\ref{eq69}).

It is particularly simple to solve Eq. (\ref{eq80}) in the binary ($n=2$) case -- see Appendix \ref{app:avalanches}.

\section{Finite Size Effects}\label{secfinite}
Let us briefly discuss the consequences of finite size effects -- namely, if we only have a finite number of $N$ agents, to what extent is the phenomenology discussed above altered?  This is important: any real experiment has finite $N$.   A more detailed analysis of finite-size effects in the $n=2$ case is present in \cite{Lucas2013} -- here we briefly discuss the extension to $n>2$.   We now return to the assumption that the graph is fully connected for all remaining computations in this paper.

Finite size effects come from fluctuations in the realization of the probability distribution $\mathrm{P}(V_1 > u_1, \ldots, V_n > u_n)$.  These lead to fluctuations in effective free energy $G$:  i.e. $G\rightarrow G+ \Delta G$, with $\Delta G$ the small fluctuations dependent on the realization of disorder.   These fluctuations then induce fluctuations in $\vec q^*$:   $\vec q
^* \rightarrow \vec q^* + \Delta \vec q$, with $\vec q^*$ the mean-field result.   We have:\begin{equation}
\vec q^* + \Delta\vec q= -\left.\frac{\partial (G+\Delta G)}{\partial \vec q}\right|_{\vec q = \vec q^*+\Delta \vec q}.
\end{equation}At leading (first) order in fluctuations we find \begin{equation}
\Delta q_i = (1-\alpha)^{-1}_{ij} \left(-\frac{\partial \Delta G}{\partial q_j}\right)_{\vec q = \vec q^*}.
\end{equation}Recall that the function $G$ is chosen so that by construction, $-\partial G/\partial q_i$ is equal to the probability that any agent prefers choice $i$, given the choices $\vec q$ of all others.   Since the intrinsic utilities $V_{i,\alpha}$ are i.i.d. random variables, we find \begin{equation}
\left(-\frac{\partial \Delta G}{\partial q_i}\right)_{\vec q = \vec q^*} = \frac{1}{N}\sum_\alpha x_{i,\alpha} - q_i^*.
\end{equation}with $x_{i,\alpha}$ i.i.d. (in $\alpha$) random variables, such that $\mathrm{P}(x_{i,\alpha}=1)=q_i^*$.  This fixes the probability distribution of $\Delta q_i$. From it, we see that finite size effects are suppressed by a $\frac1{N}$ factor. For instance, the covariance matrix is\begin{equation}
\mathrm{Cov}(\Delta q_i, \Delta q_j) = \frac{1}{N} (1-\alpha)^{-1}_{ik}(1-\alpha)^{-1}_{jl}  \mathcal{C}_{kl},
\end{equation}where \begin{equation}
\mathcal{C}_{kl} = q^*_k \delta_{kl} - q^*_k q^*_l.
\end{equation}
So long as $N\gtrsim 100$, it is therefore unlikely that this type of finite size effect alters our results in any appreciable way, unless we are close to a phase transition (where an eigenvalue of $\alpha$ tends to 1).

In addition to finite size effects, there are also finite network effects (which persist even when $N\rightarrow \infty$) -- these are consequences of finite $\langle k\rangle$.   In particular, not every node sees enough neighbors to effectively be described by mean field effects.   The most important change this induces is that the fraction of nodes in state $i$ that a node with $k$ edges sees -- denoted with $q_{i,k}$ -- fluctuates from node to node.  In the $n=2$ case, these fluctuations smooth out the energy landscape and suppress discontinuous phase transitions \cite{Lucas2013}.   We expect this smoothing phenomenon to carry over to the $n>2$ case.

\section{Oscillations in Collective Ternary Decision Making}\label{sec8}
So far in this paper, we have only discussed the case where the decision model settles to a stable fixed point.  This assumes that Eq. (\ref{eqm1}) has a stable solution.  However,  there are functions $\vec f(\vec q)$ for which there are no stable fixed points!  Thus, it may be the case (as can happen in evolutionary game theory \cite{Mobilia2010}) that our model describes persistent dynamics.   In these cases, we conclude that an energy $E$ cannot be (globally) well-defined, as otherwise the dynamical evolution of the system Eq. (\ref{gradientG}) ($\dot{\vec q} = -\nabla_f G  -\vec q$) necessarily tends towards a (local) minimum of $E$.    As first order dynamics on the real line must always tend to a fixed point, we study a model with $n=3$, the first case where persistent dynamics can arise.

A simple example of this is as follows.  Let us consider the logistic distribution $F_i(u) = [1+\mathrm{e}^{-\beta u}]^{-1}$.  We then choose \begin{subequations}\begin{align}
f_1 &= cq_2 - q_3, \\
f_2 &= cq_3 - q_1, \\
f_3 &= cq_1 - q_2.
\end{align}\end{subequations} with $c>0$.\footnote{Note that $\det(\partial f_i/\partial q_j)=0$ at all points for this model, and we can no longer define any potential $E(\vec q)$, even locally.}  Numerically, we find that the only fixed point of Eq. (\ref{gradientG}) (where $\dot{q}_i=0$) is $q_i=1/3$ for all $i$.    We determine the stability of this fixed point by standard methods \cite{strogatzbook}, and find that it is unstable so long as \begin{equation}
\beta(1-c)>8. \label{eq87}
\end{equation}As the dynamics is constrained to the simplex, we conclude by the Poincar\'e-Bendixson theorem that the dynamics tends to a limit cycle if Eq. (\ref{eq87}) is satisfied \cite{strogatzbook}.   Note that $c<1$ is required for a limit cycle to exist -- this is consistent with similar results from \cite{Mobilia2010}.

\section{Conclusion}
In conclusion, we have argued that the random field Potts model (and a wide variety of generalizations) are reasonable models for collective decision making with interacting, heterogeneous agents.   Unlike in previous works, our mean field analysis has allowed for a thorough analytic discussion of the phase diagram of the model under a variety of types of heterogeneity.   We have argued that with multiple choices, the presence of discontinuous phase transitions -- analogous to jumps and market crashes -- is an incredibly generic phenomenon.

Let us briefly discuss extensions of this work.  One interesting thing to do would be to consider the model on a graph where $n\gtrsim \langle k\rangle$.   In this case, we expect more interesting phases to emerge, where the clusters of nodes $\alpha$ whose states $i$ are undetermined percolate through the entire graph.   This should correspond with interesting dynamical phenomena and a possible emergence of glass-like physics.   The antiferromagnetic Potts model (without random fields) is equivalent to the NP-hard graph coloring problem;  the rich phase diagram of this model \cite{krzakala2} may have interesting implications for antagonistic social decision making.    It will also be interesting to consider adding ``supply-side" behavior to this model, as in \cite{Gordon2013}, which leads to the study of a ``competitive" market with interacting agents.   In particular, crucial questions to ask become whether profit-maximizing suppliers can stabilize markets against phase transitions/market crashes, the consequences of interactions on oligopolies, etc.    Finally, to the extent to which it is reasonable to consider social networks as approximately living in a two dimensional space \cite{bouchaud2d1, bouchaud2d2, fernandez}, a natural extension of our energy function -- valid over length scales much larger than the 	``lattice spacing" of the graph -- to \begin{equation}
\mathcal{E} = \int \mathrm{d}^2 x \; \left[E(\vec q(x)) + \frac{D_{ij}}{2}\nabla q_i(x)\cdot \nabla q_j(x)\right]
\end{equation}would allow us to study the relaxational dynamics of spatial patterns using an Allen-Cahn equation \cite{allencahn}.   In a SSB phase, an initial condition where different regions of space are in different minima of the energy will relax very slowly to the global minimum (where $\vec q(x)$ is $x$-independent) due to the slow dynamics of boundaries between different regions \cite{bray1, bray2, sire}; see \cite{Bouchaud2014} for similar ideas.

We are reaching an era where direct experiments \cite{Salganik2006, Lorenz2011, Moussaid2013, facebook} may be used to probe social behavior, or existing data can be analyzed from the framework of testing statistical mechanics models \cite{bouchaud2d1, bouchaud2d2, fernandez, abrams, sornette, sneppen, Lee2013}.   It is thus important to understand what are reasonable empirical tests for these models.  Signatures of the collective decision making in this paper certainly include discontinuous jumps and phase transitions and ``glassy" physics associated with a large number of equilibria.   These are both phenomena beyond the classical paradigm of economics -- an observation of the latter in particular would be strongly suggestive that this type of model is capturing qualitative behaviors of social systems.   Extending the analysis of avalanches in this paper, and comparing the statistics of avalanches on random graphs with the distribution of avalanche sizes in empirical data may also be a fruitful direction.\footnote{Anomalous heavy tails in the distribution of avalanche sizes are possible on graphs with a very heterogeneous degree distribution \cite{Lucas2013}.}   However, these phenomena are generic to disordered spin models and thus may not be helpful for ruling out any over any others.   Two phenomena which may be more specific to this model are the relationship between spontaneous ``hierarchy" formation and bimodal ``utility" distributions, and the fact that markets always spontaneously break permutation symmetry instead of jumping between two permutation symmetric points.   We hope that some of these phenomena may be experimentally confirmed in the near future.

\section*{Acknowledgements}
We would like to thank Mark Newman, Shoucheng Zhang, and anonymous referees for helpful comments.  C.H.L. is supported by a scholarship from the Agency of Science, Technology and Research of Singapore. A.L. is supported by the Smith Family Graduate Science and Engineering Fellowship.   We would both like to thank Perimeter Institute for Theoretical Physics for hospitality during the latter stages of this work.  Research at Perimeter Institute is supported by the Government of Canada through Industry Canada and by the Province of Ontario through the Ministry of Economic Development \& Innovation.

\appendix

%\section*{Acknowledgements}\addcontentsline{toc}{section}{Acknowledgements}

\section{Results for the Logistic Distribution}
\label{app:logistic}

As introduced in Sect. \ref{binarysubsub}, there exists an ansatz for single-peaked intrinsic utility functions such that $G(\vec q)$ and hence $E(\vec q)$ have an analytic closed form.  

Suppose the peak for choice $i$ is centered around $U_i=a_i$, and that each peak has a spread (variance) of $\sigma^2$ that is the same for all choices.  A convenient distribution is
\begin{equation}
F'_i(u)=\frac{\beta}{4}\mathrm{sech}^2\frac{\beta(u-a_i)}{2} 
\end{equation}
where $\sigma^2=\pi^2/3\beta^2$, with the CDF $F_i(u)$ taking the simple logistic (or Fermi-Dirac) form
\begin{equation} F_i(u)=\frac{1}{1+\mathrm{e}^{-\beta(u-a_i)}},\label{fdapp}\end{equation} 
$G$ is obtained by substituting Eq. (\ref{fdapp}) into Eq. (\ref{geq}):
\begin{eqnarray}
G&=&\lim_{R\rightarrow \infty}\int\limits^R_{-\infty} \mathrm{d}u\prod_j\frac{1}{1+\mathrm{e}^{\beta(f_i+a_i-u)}}\notag\\
&=&\lim_{R\rightarrow \infty}\frac{1}{\beta}\int\limits_{\mathrm{e}^{-\beta R}}^{\infty} \frac{\mathrm{d}z}{z}\prod_j\frac{1}{1+\alpha_jz}\notag\\
&=& R-\frac{1}{\beta }\sum_{i=1}^{n}\alpha_i^{n-1}\log\alpha_i\prod^n_{j\neq i}\frac{1}{\alpha_i-\alpha_j}\notag\\
&\rightarrow&-\frac{1}{n-1}\sum_{i=1}^{n}\frac{\partial}{\partial \beta }\left(\alpha_i^{n-1}\right)\prod^n_{j\neq i}\frac{1}{\alpha_i-\alpha_j}
\label{gfdd}
\end{eqnarray}
where $\alpha_i=\mathrm{e}^{\beta (f_i+a_i)}$. One can interpret $\mathrm{e}^{\beta a_i}$ as some kind of market "fugacity" where the offset $a_i$ takes the role of the chemical potential of choice $i$. In the last line, we have dropped an irrelevant additive constant $R$ resulting from the regularization of the divergent integral.   Care has to be taken in handling the upper limit $R$, since the integral diverges linearly with $R$. The third line can be obtained from the second line by contour integration or partial fraction expansion. 

Eq. (\ref{gfdd}) is somewhat opaque -- in particular, it looks highly singular as $\alpha_i \rightarrow \alpha_j$.   In fact, this expression is perfectly regular as $\alpha_i \rightarrow \alpha_j$ (as it has to be -- from the integral definition of $G$, there is certainly no singular behavior as $q_i\rightarrow q_j$).   To see this, let us suppose that, without loss of generality, $q_1,\ldots, q_l$ are all approaching some universal value $q_*$, and all other $q$s are distinct.  Equivalently, $\alpha_1,\ldots,\alpha_l \rightarrow \alpha_*$.   Then the singular contributions to $G$ associated with these coincident $\alpha$s appear to be the first $l$ terms, which we may approximate at leading order in this limit as \begin{equation}
G_l \approx -\frac{\log \alpha_*}{\beta} \prod_{j>l} \frac{1}{\alpha_* - \alpha_j} \sum_{j=1}^l \alpha_j^{n-1}\prod_{i\ne j} \frac{1}{\alpha_j-\alpha_i}.
\end{equation}Focusing for simplicity only on the final term (the sum from $j=1,\ldots,l$), we may re-write it as
\begin{widetext}
\begin{equation}
\sum_{j=1}^l \alpha_j^{n-1}\prod_{i\ne j} \frac{1}{\alpha_j-\alpha_i} = \dfrac{\displaystyle \sum_{j=1}^l (-1)^{j-1} \alpha_j^{n-1}\prod_{j\ne i<k} (\alpha_i-\alpha_k)}{\displaystyle \prod_{i<j\le l} (\alpha_i - \alpha_j)}.
\end{equation}
It will suffice to focus on the numerator of the above expression, and show that it is proportional to $\alpha_1-\alpha_2$ -- by permutation symmetry, it is therefore linear in all pairs $\alpha_i-\alpha_j$, and thus regular.   The terms in the sum from $j=3,\ldots,l$ are manifestly linear in $\alpha_1-\alpha_2$, so let us look at the first two terms: \begin{align}
\sum_{j=1}^2 (-1)^{j-1} \alpha_j^{n-1}\prod_{j\ne i<k} (\alpha_i-\alpha_k) &= \prod_{i,j >2} (\alpha_i-\alpha_j)\left[\alpha_1^{n-1} \prod_{j>2} (\alpha_2-\alpha_j) - \alpha_2^{n-1} \prod_{j>2} (\alpha_1-\alpha_j) \right] \notag \\
&= \prod_{i,j >2} (\alpha_i-\alpha_j) \sum_{\text{sets of }m} \prod (-\alpha_m) \left[\alpha_1^{n-1} \alpha_2^{n-2-n_m} - \alpha_2^{n-1}\alpha_1^{n-2-n_m} \right] \notag \\
&= \prod_{i,j >2} (\alpha_i-\alpha_j) \sum_{\text{sets of }m} \prod (-\alpha_m) \left[\sum_{j=0}^{n_m} \alpha_1^{n_m-j}\alpha_2^j \right] (\alpha_1-\alpha_2)(\alpha_1\alpha_2)^{n-2-n_m} 
\end{align}where in the above equations, $n_m$ is the number of $m$s in the product of $-a_m$.  We see that the first two terms are also linear in $\alpha_1-\alpha_2$, thus verifying that $G$ is regular.

\end{widetext}

In the large $\beta$ limit, there is no market heterogeneity and $G$ reduces to the very simple expression
\begin{equation}
\lim_{\beta\rightarrow \infty}=-\text{max}(f_1+a_1,...,f_n+a_n)
\label{Gzerotemp}
\end{equation}
A finite $\beta$ hence simply corresponds to a smoothening of $G$ in Eq. (\ref{Gzerotemp}), despite the seemingly complicated Eq. (\ref{gfdd}). 

An analytic solution will still exist even if $\beta=\beta_i$ are not the same for all $F_i$. That, however, calls for more than simple partial fraction expansions and the resultant expression looks less elegant.

The MF equilibrium values for $q_k$'s are given by 
\begin{eqnarray}
q_k&=&-\frac{\partial G}{\partial f_k}\notag\\
&=&-\frac{\partial}{\partial f_k}\left[\frac{-1}{\beta }\sum_{i=1}^{n}\alpha_i^{n-1}\log\alpha_i\prod^n_{j\neq i}\frac{1}{\alpha_i-\alpha_j}\right]\notag\\
&=&\alpha_k\frac{\partial}{\partial \alpha_k}\left[\sum_{i=1}^{n}\alpha_i^{n-1}\log\alpha_i\prod^n_{j\neq i}\frac{1}{\alpha_i-\alpha_j}\right]
\label{mfqq}
\end{eqnarray}
We can explicitly check that conservation of probability holds: 
\begin{eqnarray}
\sum_k^n q_k &=& \vec \alpha\cdot \frac{\partial G(\vec\alpha)}{\partial \vec \alpha}\notag\\
&=& \frac{\partial G(t\vec\alpha )}{\partial t}|_{t=1}\notag\\
&=& \frac{\partial}{\partial t}\left[G(\vec\alpha)+\log t\sum_i^n\prod_{j\neq i}^n \frac{\alpha^{n-1}_i}{\alpha_i-\alpha_j} \right]_{t=1}\notag\\
&=&\sum_i^N\prod_{j\neq i}^n \frac{\alpha^{n-1}_i}{\alpha_i-\alpha_j}\notag\\
&=&1
\label{mfq1}
\end{eqnarray}
This proof is reminiscent of Euler's theorem on homogeneous functions. Indeed, our function $G$ can be thought of as a ``partially" homogeneous function, with a part homogeneous with degree zero and a nonhomogenous logarithmic contribution.

From Eq. (\ref{mfq1}), we also see that 
\begin{equation} G(f_1+a,f_2+a,...)=G(f_1,f_2,...) +a \label{overallshift}\end{equation}
so that an overall shift in the utility functions has no physical effect. Also, scalar rescalings of the social utility $\vec f\rightarrow A\vec f$ can be absorbed in the inverse temperature $\beta$. To see how, denote $G_\beta$ and $E_\beta$ as the potentials corresponding to inverse temperature $\beta $. Then
\begin{equation} G_\beta (A\vec f) =A G_{\beta A}(\vec f)
\label{gktrans}
\end{equation}
and, if $\vec f = -\nabla_q \mathcal{H}$ (this implies that $\mathcal{H}\rightarrow A\mathcal{H}$ under a rescaling):
\begin{eqnarray} E_\beta (A\vec f) &=&G_\beta (A\vec f) + \left(\mathcal{H}-\vec q\cdot \frac{\partial \mathcal{H}}{\partial \vec q}\right)\notag\\
&=&G_\beta (A\vec f) + A\left(\mathcal{H}_{A=1}-\vec q\cdot \frac{\partial \mathcal{H}_{A=1}}{\partial \vec q}\right)\notag\\
&=& AE_{\beta A}(\vec f)
\label{Ektrans}
\end{eqnarray}
Note that Eq. (\ref{Ektrans}) holds only if $\beta$ is the same for all choices. 

\section{A detailed study of the binary ($n=2$) case}
\label{choice2}

We explore the model introduced in Sect. \ref{binarysubsub}. With the effective intrinisic utilities given by $F_i(u)=\frac{1}{1+e^{-\beta(u-a_i)}}$, Eq. \ref{gfdd} with $n=2$ reduces to 
\begin{eqnarray}
G&=& -\frac{ f_1  e^{\beta   f_1 }- f_2e^{\beta   f_2}}{\mathrm{e}^{\beta   f_1 }-\mathrm{e}^{\beta   f_2}}\notag\\
&=&  -f_2 + \frac{ f_1 - f_2}{\mathrm{e}^{\beta( f_2- f_1)}-1}
\label{gfd2}
\end{eqnarray}
Plugging in the explicit expressions $f_i=a_iq_i+b_i$ where $q_1=q$ and $q_2=1-q$, we obtain $f_2-f_1=A-Bq$, where $A=b_2+a_2-a_1$ and $B=b_1+b_2$. Hence 
\begin{eqnarray}
&&E(q)\notag\\
&=&G(q)+\mathcal{H}(q)-\sum_{q_i}q_i\frac{\partial \mathcal{H}(q)}{\partial q_i}|_{q_1+q_2=1}\notag\\
&=&G(q)+ \frac{b_1q^2+b_2(1-q)^2}{2}\notag\\
&=& -a_2-b_2(1-q)+\frac{Bq-A}{e^{\beta(A-Bq)}-1}+\frac{b_1 q^2 + b_2(1-q)^2}{2}\notag\\
&=& \frac{Bq-A}{e^{\beta(A-Bq)}-1}+\frac{b_1+b_2}{2}q^2 +\mathrm{const}.\notag\\
&=& B\left(\frac{q^2}{2}+\frac{q-C}{e^{\beta'(q-C)}-1}\right)+\mathrm{const}.
\label{E2app}
\end{eqnarray}
where $C=\frac{A}{B}$ and $\beta'=B\beta$. 

\subsection{High and low $\beta'$ limit}
In the large $\beta'$ limit, 
\begin{equation}
\lim_{\beta'\rightarrow \infty}E(q) = (C-q)\theta(q-C)+\frac{q^2}{2}
\end{equation}
which reduces to the homogeneous case result $E(q)= -\max(q-U_0,1-q)+(q^2+(1-q)^2)/2$ upon the identification $C=(1-U_0)/2$ and $\beta\rightarrow +\infty$. 

In the opposite limit of small $\beta'$,  
\begin{eqnarray}
\lim_{\beta'\rightarrow 0}E(q) &=& B\frac{q^2}{2}+\frac{1}{\beta}\left(1-\frac{\beta'}{2}(q-C)\right)+\mathrm{O}(\beta') \notag\\
&=& \frac{B}{2}\left(q-\frac{1}{2}\right)^2 +\mathrm{const}.
\end{eqnarray}
which suggest that high levels of disorder or social forces, the system assumes the maximal entropic state $q_1=q_2=\frac{1}{2}$, \emph{independently} of any intrinsic utility.

\subsection{Conditions for bistability}
One can obtain necessary conditions for bistability, i.e. having two minima by expanding the potential about $q=C$ to quadratic order: $E\approx \left(\frac{1}{2}-\frac{\beta'}{12}\right)\left(q-\frac{3-\beta' C}{6-\beta'}\right)^2+\mathrm{const}$. Since we require an unstable equilibrium in the middle, $\frac{1}{2}-\frac{\beta'}{12}<0$ or $\beta'=\beta B>6$. Also, the unstable equilibrium must occur at $0<q<1$, so $\beta'C=\beta A >3$. Both conditions require that $\beta$ is sufficiently large, i.e. that the agents are sufficiently homogeneous in their intrinsic utilities. This is a sensible precondition for any sudden market transition (crash). Furthermore, $\beta A$ must also be large enough, or there will not be sufficient utility differential between the choices to drive the crash. Ultimately, the social effect $\beta'=\beta B$ must also be large enough to create two basins in the potential, so that a bifurcation can occur. In fact, there \emph{must} be two basins of attraction in the limit of large positive social effect, as evidenced from the discontinuity in the large $\beta'$ limit of $E$:  $(C-q)\Theta(q-C)+q^2/2$.% when $\frac{A}{B}>0$. 

One can obtain the conditions for bistability to any degree of accuracy through the graphical solution described next.

\subsection{Location of the fixed points}
We now solve for the fixed points explicitly via $q_i=-\partial G/\partial f_i$. From Eq. (\ref{gfd2}),
\begin{eqnarray}
G%&=& -\frac{f_1 e^{\beta f_1}-f_2e^{\beta f_2}}{e^{\beta f_1}-e^{\beta f_2}}\notag\\
&=&-\frac{\partial}{\partial \beta }\log(\mathrm{e}^{\beta f_1}-\mathrm{e}^{\beta f_2})
\end{eqnarray}
Switching orders of differentiation, it follows that that 
\begin{equation}
q=q_1= Q(x)=\frac{1}{1-\mathrm{e}^x}+\frac{x\mathrm{e}^x}{(1-\mathrm{e}^x)^2}
\end{equation}
where $x=\beta (f_2-f_1)$. This equation can be solved self-consistently. In particular, it depends only the \emph{difference of the utilities} of the two choices. This makes sense: with the spreads in the intrinsic utilies being equal, there is no "intrinsic" reluctance to switch choices and the fixed points thus depend only on the difference in the utilties.

Now let us solve for the specific model in Sect. \ref{binarysubsub} explicitly. With $f_1=a_1+b_1q$ and $f_2=a_2+b_2(1-q)$,
\begin{equation} 
x=\beta (f_2-f_1)=\beta (A - Bq)  
\end{equation} 
where $A=b_2+a_2-a_1$ and $B=b_1+b_2$. Hence the the solution for the fixed points can be found from the intersection of $q=Q(x)$ and the linear graph $q=(A-x/\beta)/B$. 

\subsubsection{Analysis and physical interpretation of the fixed points}
Analysis of the possible behaviors is simple, because only the straight line has parametric dependence.  We see that the x-intercept occurs at $x=\beta A$ and the slope is $-1/B\beta$.   Only one solution is possible if $\beta B=\beta (b_1+b_2)<6$, since that is the maximum downward slope of $Q(x)$. This is just saying that we need a minimal amount of social interactions ($q$ dependence) before phase transitions are possible, no matter how skewed the intrinsic utilities are.

\begin{figure}
\begin{minipage}{\linewidth}
\centering
\includegraphics[width=.7\linewidth]{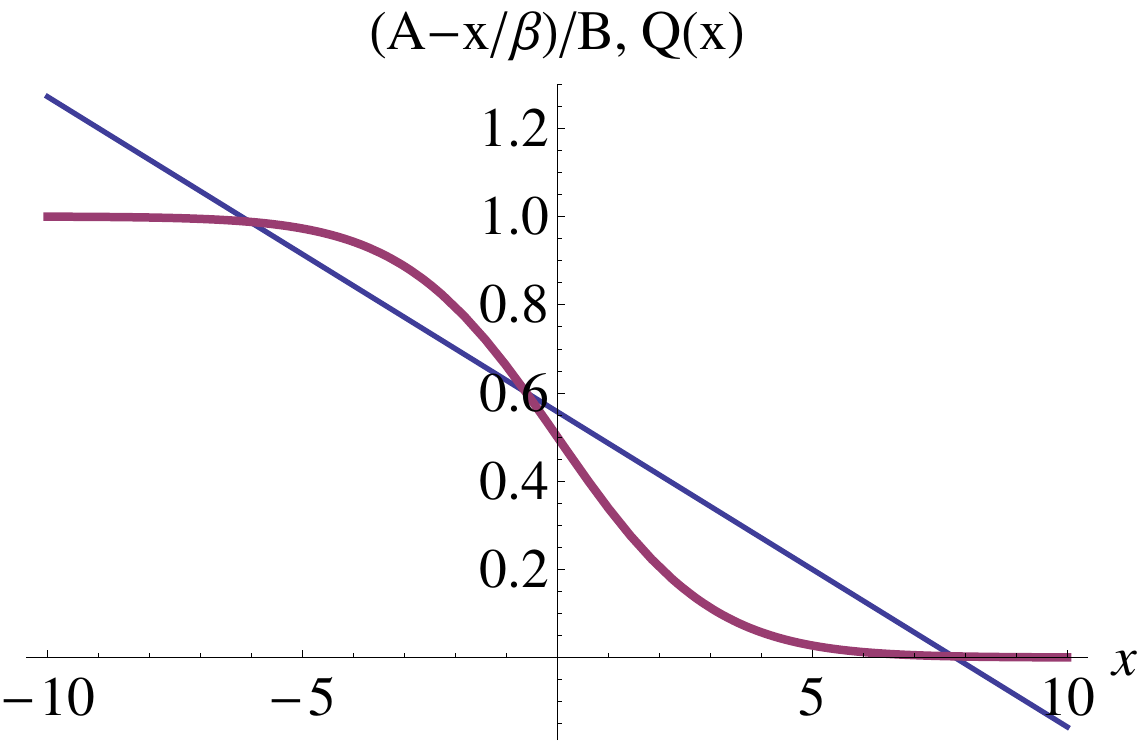}
\end{minipage}
\caption{Plots of $y=Q(x)$ (black) and $y=B^{-1}(A-x/\beta)$ (blue). Here $\beta =10$, $A=0.78$ and $B=1.4$. The y-axis is magnified ten times for clarity. The left and right solutions are stable, while the middle one is unstable.}
\end{figure}

If $\beta B$ is very large, the line will intersect the curve very far away. Hence the stable solutions of $q_1$ will be either near $1$ or $0$:  very strong social interactions result in an all-or-nothing scenario.  

From the graph, we need $\beta A=\beta (a_2-a_1)>3$ for 3 solutions to be possible. Thus, a phase transition can only happen if the intrinsic utility of choice 2 is sufficiently larger than that of 1, no matter how big or small are the social interactions. This should be true in all $n=2$ models with linear utilities.   

Finally, if we start from a large $x$-intercept $\beta A$ and increase $\beta B$ from a negative value, the straight line will rotate about the intercept anticlockwise until it suddenly has 2 new solutions with large $q_1$ (though there will be only 1 solution exactly at a critical point). This is just saying that with sufficient initial intrinsic disutility of choice 1, a phase transition must occur as social interactions are introduced.

\section{More General Gradient Flow}\label{appgradflow}

Gradient flow dynamics exist in the general case where where $\vec f = -\nabla_q \mathcal{H}$ i.e. that the MF social utility is derivable from a microscopic Hamiltonian, \emph{and} $\partial f_i / \partial q_j$ is either positive definite or negative definite. Intuitively, the requirement for positive or negative definiteness avoids degenerate points where $ \frac{\partial f_i}{\partial q_j}=0$, which generically imply a noninjective utility function $\vec f(\vec q)$. An unique energy surface can only be defined if there is a one to one correspondence between  configurations $\vec q$ and their utilities $\vec f$.

For the gradient flow, we consider a change of variables to $\vec \gamma (\vec q)$, with $\partial \gamma_i/\partial q_j$ positive definite. Multiplying  Eq. (\ref{gradientG}) with $\partial \gamma_k/\partial q_i$, \begin{equation}
-\frac{\partial \gamma_k}{\partial q_i}\left(\frac{\partial G}{\partial f_i} - q_i\right) = \frac{\partial \gamma_k}{\partial q_i}\dot{q}_i = \dot{\gamma}_k
\end{equation}
If we choose \begin{equation}
\frac{\partial \gamma_k}{\partial q_i}\frac{\partial \gamma_k}{\partial q_j} = \pm \frac{\partial f_i}{\partial q_j}  \label{eqb2}
\end{equation}where the $\pm$ sign is chosen if $\partial f_i/\partial q_j$ is positive (negative) definite, some matrix manipulations similar to Eq. (\ref{eq14}) reveal that \begin{eqnarray}
\dot{\gamma}_k &=& \mp \frac{\partial}{\partial \gamma_k} \left(G + \mathcal{H} + q_i f_i\right)\notag\\
&=& \mp\frac{\partial E}{\partial \gamma_k}.
\label{gradientflowgen}
\end{eqnarray}

Therefore, we see that gradient flow dynamics is consistent for rather general class of $\vec f (\vec q)$. 

\subsection{Linear Utilities}
One example is the particularly simple case where \begin{equation}
\vec f  = A \vec q
\end{equation}with $A$ a matrix.  In the main text, we have focused on the case where $A$ is the identity matrix. %Hence the matrix $A$ allows for differing marginal social utilities for the various choices. 
Here, we allow the social utility $f_i$ of choice $i$ to depend on the $q_j$ for all the various choices $j$. For instance, the social utility of a certain social medium website may also depend positively on the popularity of certain sister sites, and negatively on that of rival sites. Since $A$ must be positive definite, $A$ is symmetric. This means that the utility functions are \emph{reciprocal}: If $f_i$ depends on the proportion of agents subscribed to choice $j$ via $f_i=\lambda q_j+...$, then $f_j=\lambda q_i+...$ too.% We do not need to include a constant offset in this definition of $\vec f$, since that can always be absorbed into the intrinsic utility function functions $\vec F$. 

Write \begin{equation}
A = S^{\mathrm{T}}DS
\end{equation}with $D$ a diagonal matrix with positive eigenvalues, and $S$ an orthogonal matrix.   We then define \begin{equation}
\vec \gamma = D^{1/2}S \vec q.
\end{equation}It is easy to check that this satisfies Eq. (\ref{eqb2}).  Since $H-\vec q^T \cdot\frac{\partial H}{\partial \vec q}=\frac{1}{2}\vec q^TA\vec q$, we obtain \begin{equation}
E(\vec \gamma) = \frac{1}{2}|\vec \gamma|^2+ G\left(\vec f \right) \label{Elinear}
\end{equation}where $\vec f = S^{\mathrm{T}}D^{1/2}\vec \gamma$. We still retain the same quadratic term as in the $\vec f=\vec q$ case, but with a $G$ that depends on an argument rotated and rescaled by $S^{\mathrm{T}}D^{1/2}$. Note that the configuration space simplex is also rotated and rescaled, but in the opposite way $D^{1/2}S$. 
%When $A$ is a multiple of the identity and the intrinsic utilities are all of the same logistic type, Eq. \ref{Elinear} reduces to Eq. \ref{Ektrans}.

\subsection{Nonlinear Noncooperative Utilities}

Nonlinear utility functions allow for varying levels of marginal utilities at different stages of market domination, and can thus represent real scenarios more realistically. Consider the simplest case where each $f_i$ is an arbitrary monotonic function of $q_i$ only, i.e. the utilities of the different choices decouple.  The matrix $\partial f_i/\partial q_j$ is then diagonal and we can simply find $\gamma$: 
\begin{equation}
\gamma_i = \int\sqrt{\frac{\mathrm{d}f_i}{\mathrm{d}q_i}}\mathrm{d}q_i.
\label{diagonalgamma}
\end{equation} 

\subsubsection{Power-law utilities}  
As the simplest example, consider the case where \begin{equation}
f_i = g_i q_i^{\eta_i}.
\end{equation}Then we find that \begin{equation}
E = \sum_i \frac{\eta_i+1}{4}\gamma_i^2  + G\left(\vec f(\vec \gamma)\right)
\end{equation}where \begin{equation}
\gamma_i = \frac{2\sqrt{\eta_ig_i}}{\eta_i+1}q_i^{(\eta_i+1)/2}.
\end{equation}
and 
\begin{equation}
f_i(\gamma_i) = g_i\left(\frac{\eta_i+1}{2\sqrt{\eta_ig_i}}\right)^{2\eta_i/(\eta_i+1)}\gamma_i^{2\eta_i/(\eta_i+1)}.
\end{equation}

\subsubsection{Logarithmic utilities}
\label{app:nonlinearlog}

Now consider the logarithmic utility function
\begin{equation} f_i=g_i\log (q_i+\delta) \end{equation}
with $\delta$ a small regularizing constant. Here, the marginal utility approximately proportional to the \emph{fractional} change of the market share $q_i$. From Eq. \ref{diagonalgamma}, we obtain $\gamma_i=2\sqrt{g_i(q_i+\delta)}$. This leads us to
\begin{eqnarray}
E&=& \sum_i \frac{\gamma_i^2}{4}+(g_i\delta)\left(\log \frac{\gamma_i^2}{4g_I}-1\right)+G\left(\vec f(\vec \gamma)\right)\notag\\
&=& \sum_i g_iq_i +g_i\delta\log(q_i+\delta)+ G(\vec f)\notag\\
&\approx& \sum_i g_iq_i +G(\vec f)
\end{eqnarray}
which bears superficial similarity with the above power-law case with $\eta_i=0$. In the last line, we have dropped the logarithmic term, which tends towards zero as $\delta\rightarrow 0$. This energy potential is analyzed graphically in Fig. \ref{fig:nonlinearlog}. Notice that in all of the above cases, $E$ always contain quadratic contributions in $\gamma_i$.

\section{More on Ternary Decision Making}
\label{app:permsymm}

\subsection{The Permutation Symmetric Logistic Case}
Here we provide additional details on logistic permutation symmetric ternary decision making.  As in the main text, we work in barycentric coordinates;  we will also complexify them as in Eq. (\ref{baryeq}), and set $\bar z = x-\mathrm{i}y$. By a brute-force expansion of Eq. (\ref{gfdd}), the energy $E$ is 
\begin{eqnarray}
E&\approx& \left(\frac{1}{3}-\frac{\beta}{12}\right)|z|^2-\frac{\beta^2}{180}|z|^3\cos 3\theta+\frac{\beta^3}{720}|z|^4\notag\\
&&+6\times 10^{-6}\beta^5 |z|^6 + \cdots
\label{phase3}
\end{eqnarray}
which is very accurate for $\beta \gtrsim 1$, away from the simplex corners. Recall that \begin{equation}
|z|^2 = \frac{3}{2}\sum_{i=1}^3 \left(q_i-\frac{1}{3}\right)^2
\end{equation}denotes the distance from the permutation symmetric fixed point.

We see that the permutation-symmetric point is stable for $\beta<4$, unstable for $\beta>4$ and marginally unstable (a monkey saddle) for $\beta=4$. This makes physical sense: For small $\beta$ social interactions are suppressed by market heterogeneity and SSB should not occur. For large $\beta$, social interactions distort individual preferences, and we will expect that once one of three roughly equally intrinsically desirable products have some lead, it will continue to dominate.

\subsection{Conformal Symmetry}

%We have previously described the general approach of deriving the  potential $E$ for any social utility function $\vec f$, and explicitly showed how to construct that for decoupled nonlinear utilities. 

Since the $n=3$ simplex is planar, it possesses a conformal group (transformations which distort space but locally preserve angles) with an infinite number of generators. This fact can be exploited to easily write down the path swept out by $\vec q(t)$ near the critical point when $\vec f = \vec q$.   On the simplex, a straightforward calculation shows that \begin{equation}
\dot{x} = -\frac{3}{2}\frac{\partial E}{\partial x}, \;\;\; \dot{y} =- \frac{3}{2}\frac{\partial E}{\partial y}.
\end{equation}
Near a phase transition, we find from Eq. (\ref{poisson}) that the charge density $\rho=4\partial_z\partial_{\bar z}E\approx 0$.  Hence we can approximate $E$ as the real part of a holomorphic function $w(z)$: $E=\mathrm{Re}(w(z))$. From the Cauchy-Riemann equations, we obtain that the trajectory of $\vec q(t)$ to be along level curves of $\mathrm{Im}(w)$, i.e. normal to the level curves of $E=\mathrm{Re}(w)$.

As an example, take the abovementioned permutation symmetric model at $\beta=4$. The trajectories are given by 
\begin{align}
&\mathrm{constant} = -\frac{\beta'^2}{360}\mathrm{Im}\left(z^3\right) = \frac{\beta^2}{180}(y^3-3xy)\notag\\
&\;\;\;\;\;=\frac{\sqrt{3}\beta^2}{480}(q_1-q_2)((q_1-q_2)^2-6q_3+2)
\end{align} 
This follows from the term in $E$ containing $  z^3 + \bar z^3 \propto \mathrm{Re}(z^3)$.   This $E$ corresponds to an electrostatic potential which can be realized between a $60^\circ$ wedge whose outer edges are conductors held at different potentials.

\section{Avalanches and Complexity with $n=2$}
\label{app:avalanches}
In a previous paper \cite{Lucas2013} we used a similar formalism to derive many results for $n=2$ in a simpler way.  For pedagogy, we will explain how to use the formalism of this paper to derive our old results concerning avalanches and the complexity phenomenon, which are the trickiest to re-derive.    The key point to note is that $\alpha_{ij}$ is a symmetric $2\times 2$ matrix which has two constraints that the sums on rows/columns vanishes.    There is a unique way to write this matrix:  \begin{equation}
\alpha = \left(\begin{array}{cc} \alpha_0/2 &\ -\alpha_0/2 \\ -\alpha_0/2 &\ \alpha_0/2 \end{array}\right).
\end{equation}where $\alpha_0\ge 0$ is a constant.    This looks like a matrix describing a permutation symmetric model, although the underlying model needs no such symmetry.   The eigenvalues of $\alpha$ are 0 (corresponding to $\delta q_1 = \delta q_2$) and $\alpha_0$ (corresponding to $\delta q_1=-\delta q_2$).

Let us begin by determining the expected number of agents who will change state during an avalanche, where a node changes state from 2 to 1.  In this case, denoting with $X_i$ the expected number of agents who will change, we conclude that $X_1=-X_2$: \begin{equation}
\left(\begin{array}{c} X_1 \\ -X_1 \end{array}\right) = (1-\alpha)^{-1}\left(\begin{array}{c} 1 \\ -1 \end{array}\right) = \frac{1}{1-\alpha_0} \left(\begin{array}{c} 1 \\ -1 \end{array}\right),
\end{equation}where we exploit the fact that this vector is an eigenvector of $\alpha$.   Indeed, $\alpha_0$ corresponds to what was simply denoted $\alpha$ in \cite{Lucas2013} -- the probability that a node will flip its binary state in an avalanche.

It is also worth stressing that there is a dramatic simplification in the study of avalanches (and complexity), because all nodes in an avalanche will flip to the same state.  In fact, Eq. (\ref{eq80}) boils down to a simple algebraic equation, as the vector $\xi_{kl}$ has only one component -- $\xi_{12}$.   We find that \begin{equation}
\left(\frac{2}{\alpha_0}-2\right)\xi_{12}=1,
\end{equation}which allows us to conclude, using Eq. (\ref{eq70}), that \begin{equation}
\mathrm{P}(1,2) = \frac{\alpha_0}{\langle k\rangle} \times 2\xi_{12} = \frac{\alpha_0^2}{(1-\alpha_0)\langle k\rangle},
\end{equation}which agrees with the result derived from Thouless-Anderson-Palmer equations in \cite{Lucas2013}.

\section{An Introduction to SSB}\label{ssbintro}
Let us provide a brief reminder of what spontaneous symmetry breaking (SSB) is.    We say that our permutation symmetric models have $\mathrm{S}_n$ (permutation) symmetry because, for any element $\sigma\in \mathrm{S}_n$ ($\sigma$ is a bijective map from the set $\lbrace 1, \ldots, n\rbrace$ to itself), the energy $E(q_1,\ldots, q_n) = E(q_{\sigma(1)},\ldots, q_{\sigma(n)})$.   Thus, a global re-numbering of the choices does not alter the energy landscape.   However, a typical point in the simplex \emph{breaks} this symmetry:  in general, $(q_1,\ldots q_n) \ne (q_{\sigma(1)}, \ldots, q_{\sigma(n)})$.  Only when $q_i=1/n$ for every $i$, is the full symmetry group present.   This is called the permutation symmetric point.    We are interested in the cases where $E$ has $\mathrm{S}_n$ symmetry, but $\vec q$ at a minimum of $E$ does not have $\mathrm{S}_n$ symmetry -- these phases are called symmetry broken phases (we will often say SSB phase for shorthand).

 It may turn out, however, that some subgroup of $\mathrm{S}_n$ is preserved.   For example, if $q_1=q_2 = \cdots = q_{n-1}\ne q_n$, then so long as $\sigma(n)=n$, the permutation leaves the value of $\vec q$ unchanged.   This value of $\vec q$ preserves permutation symmetry among $n-1$ choices, so it has $\mathrm{S}_{n-1}$ symmetry.   Alternatively, it may so happen that $q_1 = q_2 = \cdots = q_p \ne q_{p+1} = \cdots = q_n$.   Because here any permutation $\sigma$ that only swaps the first $p$ elements among themselves, and the last $n-p$ elements among themselves, is allowed, we say that the symmetry of this state is $\mathrm{S}_p \times \mathrm{S}_{n-p}$ -- we have two distinct permutation symmetries.
 
 This phenomenon is incredibly important.   The fact that $E$ has $\mathrm{S}_n$ symmetry in a permutation symmetric model means that \emph{every choice is identical on average}.   SSB phases nonetheless pick out certain choices as ``better" than others -- some $q$s are higher than others.    The occurence of SSB is a signature that social interactions are playing a crucial role.   
 
 For a realistic social system, SSB may be ``hard to observe" because the system may very well not have any symmetry in the first place.  There are two reasons why SSB is nonetheless important.   Firstly, the $\mathrm{S}_n$ symmetry may be approximately present, and social interactions overwhelm the small amount of intrinsic symmetry breaking.  Secondly, SSB phases in our toy models are toy examples of phase transitions and social ``herding" phenomenon caused entirely by interactions, which probably \emph{do} play a role in the real world.  Thus, understanding what classes/instances of models are sensitive to SSB provides us with insight into the (dramatic) effects of interactions in collective decision making, in instances where (because of symmetry) our analytic abilities are much greater.

\section{Stability of the Permutation Symmetric Point in the Large $n$ Limit}\label{largenapp}
In this appendix we briefly discuss the stability of the permutation symmetric fixed point as $n$ increases.   This provides a simple answer to the question of whether more choices  enhances or suppresses SSB instability.

Let us begin by assuming that $F(u)$ is given by a logistic distribution Eq. (\ref{fdapp}) where $\beta$ is fixed, and that $\vec f = \vec q$. Straightforward integrals give us that \begin{equation}
\alpha_{11} = -\int \mathrm{d}u\; F^{\prime\prime} F^{n-1}= \frac{\beta}{n}\frac{n-1}{n+1}.
\end{equation} for the logistic distribution.   We conclude from Eq. (\ref{alpha11}) that the permutation symmetric fixed point is stable when \begin{equation}
\beta<n+1.  \label{eqd2}
\end{equation}The more choices that are given, the harder it is for herding effects to take over, and for the permutation symmetry to be broken.    

In fact, this is nearly a generic result.   To see this, note that by deriving $b$ (recall that when $b>1$, the permutation symmetric fixed point is unstable) from $\alpha_{12}$, it is straightforward to derive that \begin{equation}
\frac{b(n)}{n} = \int \mathrm{d}u F^{\prime 2} F^{n-2} >  \int \mathrm{d}u F^{\prime 2} F^{n-1} = \frac{b(n+1)}{n+1}. 
\end{equation}We see that, up to the ``mild" difference between $n$ and $n+1$, we have $b(n)\gtrsim b(n+1)$.   No ``reasonable" distribution $F$ we have ever checked has $b(n) \le b(n+1)$ for any $n$, although we have not been able to rule it out.   A heuristic explanation for this is the following:  if there are more choices, then it is more likely that people find a choice that they feel strongly about.  Thus, adding more choices would stabilize the permutation symmetric fixed point.

However, there is a rather straightforward scaling limit in which increasing the number of choices $n$ does not alter stability:  suppose that \begin{equation}
\beta = \beta_0 n
\end{equation}Here we are still studying the logistic distribution.  One might argue that this is a more physical limit, in the sense that as $n\rightarrow \infty$, there is an instability whenever \begin{equation}
\beta_0 < 1,
\end{equation}independently of $n$.    This is one mechanism for allowing an instability to persist with many choices.   

This is rather ad hoc, for the simple reason that different distributions require different scaling limits to preserve the stability of the symmetric point as $n\rightarrow \infty$.   A particularly interesting example is the case where $F(u)$ corresponds to the uniform distribution: \begin{equation}
F(u) = \left\lbrace\begin{array}{ll} 0 &\ u<0 \\ u/u_0 &\  0<u<u_0 \\ 1 &\ u>u_0 \end{array}\right..   \label{uniformdist}
\end{equation}Assume that $u_0$ does not change with $n$.  One can check rather easily for this distribution, from the definition of $\alpha_{ij}$, that \begin{equation}
\alpha_{11} = -\int\frac{\mathrm{d}u}{u_0} \left(\delta(u)-\delta(u-u_0)\right) F(u)^{n-1} = \frac{1}{u_0}, \label{unialpha11}
\end{equation}which implies that\begin{equation}
b(n) = \frac{n}{n-1} \frac{1}{u_0}.
\end{equation}So we see that whenever $u_0<1$, there is an instability in the $n\rightarrow \infty$ limit.   This suggests a rather more robust mechanism for preserving instabilities of the permutation symmetric point in the $n\rightarrow \infty$ limit.

\section{Details of the Landau Theory Caclulation}\label{landauapp}
We provide the computational details of the Landau theory calculation, whose conclusion was summarized in the main text.   

\subsection{Form of $E$}

Our starting point is to verify that Eq. (\ref{landaueq}) is the most general possible energy $E$ up to $\mathrm{O}(\delta q_i^4)$ consistent with permutation symmetry, and the simplex constraint that $\sum \delta q_i = 0$.    A linear term cannot be included in $E$ since the only possible choice consistent with the symmetry is proportional to the simplex constraint above.   The only quadratic terms allowed are $\sum \delta q_i^2$ and $\sum_{i\ne j} \delta q_i \delta q_j$ -- again, using the simplex constraint, these turn out to be proportional, so we simply choose the first one as it is manifestly positive.   The three cubic terms are given by sums over $\sum_i \delta q_i^3$, $\sum_{j\ne i}\delta q_i^2\delta q_j$, or $\sum_{i\ne j\ne k}\delta q_i\delta q_j\delta q_k$.   The first two terms are proportional for an analogous reason to before; we then use the fact that \begin{equation}
\left(\sum_i \delta q_i\right)^3 = \sum_i \delta q_i^3 + 3 \sum_{j\ne i} \delta q_i^2\delta q_j + 6\sum_{i\ne j\ne k} \delta q_i\delta q_j\delta q_k
\end{equation}vanishes to conclude that there is only a unique independent cubic term, which we take to be $\sum_i \delta q_i^3$.   Identical arguments to these lead us to conclude that at quartic order, there are two possible terms, as they are given in Eq. (\ref{landaueq}).

\subsection{The Minima of $E$}   

The assumption that there is a continuous phase transition requires that $E$ be bounded from below.  This is satisfied if $\psi>0$ and \begin{equation}
\frac{\omega}{\psi} > - \frac{n^2-3n+3}{(n-1)n}
\end{equation} or if $\psi<0$ and \begin{equation}
\frac{\omega}{|\psi|} > \left\lbrace\begin{array}{ll} n(n^2-1)/(n^2+3) &\  n\text{ odd} \\ n &\ n\text{ even} \end{array}\right..
\end{equation}

The most efficient way to find the minima of $E$ is as follows.  Let us begin by fixing $\delta q_4,\ldots, \delta q_n$.   We will show that given arbitrary $\delta q_4,\ldots, \delta q_n$,  at the local minimum of $E$ among $\delta q_1,\ldots, \delta q_3$, two of these must be equal.    Since any global minimum of $E$ must also correspond to a minimum of $E$ when constrained to this subspace, we conclude that it is impossible to have any three $i,j,k$ for which $\delta q_i \ne \delta q_j \ne \delta q_k$, in the minimum of Eq. (\ref{landaueq}).

Let us now prove the claim.  The simplex constraint implies that the sum $\delta q_1+\delta q_2+\delta q_3 \equiv 3\eta$ is also fixed.   Let us also define \begin{subequations} \label{eqe4}\begin{align}
\delta q_1 &= \eta + r\cos\theta, \\
\delta q_2 &= \eta + r\cos\left(\theta+\frac{2\pi}{3}\right), \\
\delta q_3 &= \eta + r\cos\left(\theta+\frac{4\pi}{3}\right).
\end{align}\end{subequations}Then the Landau energy is, up to a constant factor $E_0$: \begin{equation}
E = E_0 - \frac{3\zeta }{2} r^2 + \frac{9\theta + 18\psi}{8}r^4 - \left(\frac{3\xi}{4}-3\eta \omega \right) r^3\cos(3\theta).
\end{equation}The minimum of $E$ on this two dimensional subspace corresponds to a point where two of the 3 $\epsilon$s in question equal each other.   Since any global minimum of $E$ must also correspond to a minimum of $E$ when constrained to this subspace, we conclude that it is impossible to have any three $i,j,k$ for which $\delta q_i \ne \delta q_j \ne \delta q_k$, in the minimum of Eq. (\ref{landaueq}).

We conclude that the symmetry of the resulting minimum, when $\zeta >0$, must preserve a $\mathrm{S}_p\times \mathrm{S}_{n-p}$ subgroup of $\mathrm{S}_n$: i.e., \begin{equation}
\delta q_i = \left\lbrace\begin{array}{ll} \epsilon/p &\ i=1,\ldots, p\\ -\epsilon/(n-p) &\ i=p+1,\ldots,n\end{array}\right..  \label{eqe6}
\end{equation}
In fact, as the transition is discontinuous (as we will see shortly), to understand the SSB transition it will suffice to study the location of minima at $\zeta=0$.   In this case, plugging Eq. (\ref{eqe6}) into our energy ansatz, we find \begin{widetext} \begin{equation}
E = -\xi \epsilon^3 \left(\frac{1}{p^2}-\frac{1}{(n-p)^2}\right) + \epsilon^4 \left(\omega \left(\frac{1}{p^3}+\frac{1}{(n-p)^3}\right) + \psi \left(\frac{1}{p}+\frac{1}{n-p}\right)^2\right),
\end{equation}which has a minimum at energy \begin{equation}
E_{p,n} = -\frac{3^3 \xi^4}{4^4}  \left(\frac{1}{p^2}-\frac{1}{(n-p)^2}\right)^4\left(\omega \left(\frac{1}{p^3}+\frac{1}{(n-p)^3}\right) + \psi \left(\frac{1}{p}+\frac{1}{n-p}\right)^2\right)^{-3}.
\end{equation}
\end{widetext}
The fact that even at $\zeta =0$, $E_{p,n}<0$ guarantees that the transition is discontinuous.  By differentiating $E_{p,n}$ with $p$, we can find the optimal value of $p$.   It is clear $E_{p,n} = E_{n-p,n}$, we focus on $p<n/2$.   We find that $E_{p,n}$ is decreasing for $p<p^*$ and increasing for $p>p^*$, with \begin{equation}
p^* = \frac{n}{2}\left(1-\sqrt{\frac{\omega+n\psi}{3\omega + n\psi}}\right). \label{eqe9}
\end{equation}
These results are universal and hold for arbitrary permutation symmetric models which have an energy function $E$.

\subsection{Cooperative Decision Making}
So far, our discussion has been entirely based on assumptions of permutation symmetry alone -- we have not added any specific input about our decision model.   Let us consider $\vec f = \vec q$.   Then all cubic and quartic contributions come from derivatives of $G$.  Define \begin{equation}
\kappa_{ijk} \equiv \frac{\partial^3 G}{\partial q_i \partial q_j \partial q_k}, \;\; \; \kappa_{ijkl} \equiv \frac{\partial^4 G}{\partial q_i \partial q_j \partial q_k \partial q_l}.
\end{equation}
In what follows, we will denote with $\kappa_{112}$ the value of any coefficient of $\kappa_{ijk}$ with two indices equal, and the third index different; all other $\kappa$ coefficients follow analogous definitions.   By positivity of $F$ and $F^\prime$, we see that $\kappa_{123}<0$ and  $\kappa_{1234}>0$.   Using constraints that $\sum_k \kappa_{ijk}=0$, we conclude that $\kappa_{112}>0$ and $\kappa_{111} \sim -\xi < 0$.   Analogous to what we found for the $\alpha_{ij}$ matrix, $\kappa_{111}$ is not directly proportional to $\xi$, because the way we write out $G$, there will be terms proportional to $\delta q_i \delta q_j \delta q_k$, e.g.   Accounting for this properly, and using that $2\kappa_{123}  = -(n-2)\kappa_{112}$, and $\kappa_{111} = -(n-1)\kappa_{112}$, we find that, for $n>2$: \begin{equation}
\xi = \frac{|\kappa_{111}|}{6} \frac{(n-4)(n-2)+3}{(n-1)(n-2)} > 0.
\end{equation} 
A similar calculation for the quartic case reveals that \begin{subequations}\begin{align}
\psi &= \frac{\kappa_{1122}}{8} + \frac{2n-3}{12}\kappa_{1234} >0, \\
\omega &= \frac{n}{24}\kappa_{1122} - \frac{n^3-2n^2+3n-9}{72}\kappa_{1234}.
\end{align}\end{subequations}Note that by construction, $\kappa_{1122},\kappa_{1234}>0$.  We conclude that $\psi>0$.  If $\theta<0$, Eq. (\ref{eqe9}) implies that the smallest SSB phase has $\mathrm{S}_{n-1}$ symmetry.\footnote{This constraint can be slightly relaxed.   We simply need that $p^*<2$, and that $E_{1,n}<E_{2,n}$.}   

In general, for unimodal $F(u)$ distributions, we find numerically $\omega<0$.   We have found distributions for which $\omega>0$ (e.g. Eq. (\ref{bimodalfigeq}) with $n=4,5$, $u_0\gtrsim 100$, $\beta=1$, $p=0.5$), yet we have found in these cases $p^*<1$.

In conclusion, we cannot rule out in principle that there exist permutation symmetric models which directly undergo a discontinuous phase transition from the $\mathrm{S}_n$ symmetric point to a $\mathrm{S}_p \times \mathrm{S}_{n-p}$ fixed point with $p>1$, though we have constrained it to this general form.   For all practical examples we have found, however, $p=1$, even in the bimodal case.   This suggests that the dominant pattern of symmetry breaking is simply that one choice becomes more popular, with all others remaining equal.  Of course, we have shown that far away from the $\zeta = 0 $ transition, bimodal distributions can lead to further patterns of symmetry breaking -- the Landau theory description is only valid close to the permutation symmetric point.

\section{Permutation Symmetry Breaking in the Logistic Model} \label{appendixg}
The Landau theory arguments are extremely generic, but we are unable to completely prove, even for a unimodal distribution, that $p^*$ is small enough so that permutation symmetry in a model with $n$ choices is broken only from $\mathrm{S}_n$ to $\mathrm{S}_{n-1}$.   In this appendix, we demonstrate that near the SSB phase transition, the SSB pattern in logistic models is indeed $\mathrm{S}_n$ to $\mathrm{S}_{n-1}$.  Logistic distributions are toy models of generic unimodal distributions, as detailed in Appendix \ref{app:logistic}.  Since deep in the SSB phase ($\beta \gg n+1$) logistic models are well described by Eq. (\ref{Gzerotemp}), this provides a rather complete justification of our claim that this model never breaks permutation symmetry beyond $\mathrm{S}_{n-1}$.

\begin{figure}
\centering
\includegraphics[scale=0.8]{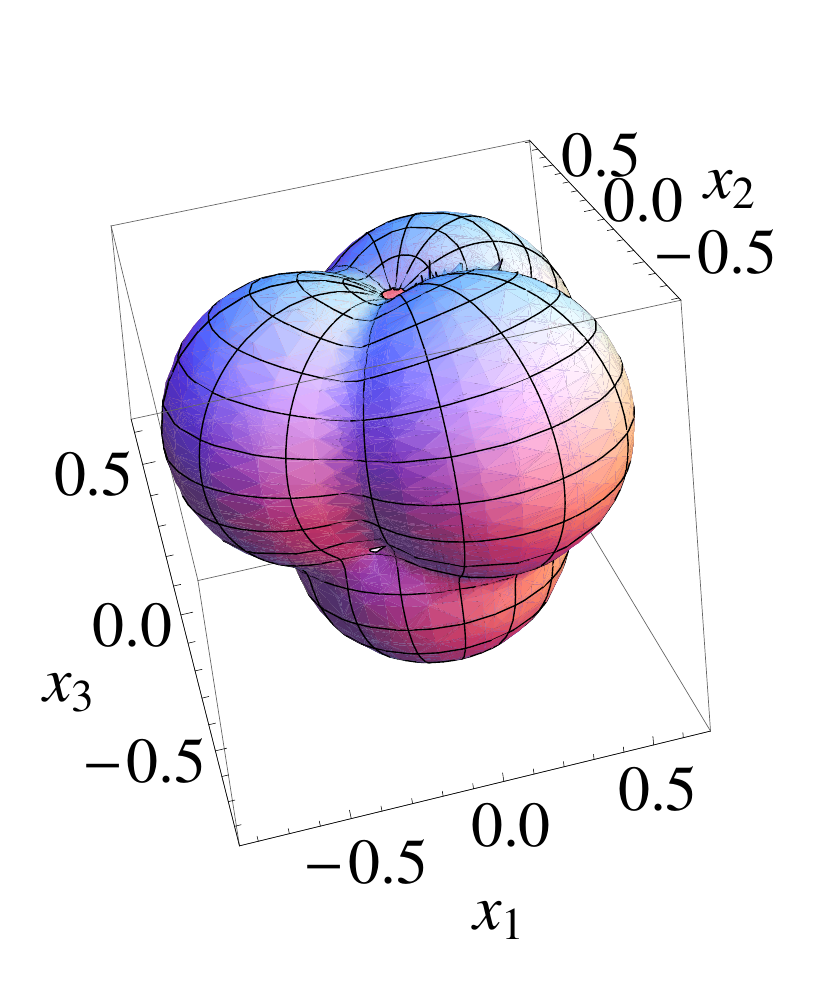}
\caption{ The profile of $G$ for $n=4$, $\beta=40$. The distance from the origin represents the magnitude of $G$ at a chosen $r=0.8$. $\mathrm{S}_3$ symmetry is evident from the ``tetrahedral" lobes.} %\red{This figure can convince the reader that shapes other than $S_{n-1}$ look improbable.} Middle,Right) Fourier coefficent magnitude $f_3(\varphi_1)$ and $f_6(\varphi_1)$ for $0.6<r<1$ and $k=4$ and   $0<\varphi_1<\pi$. Note that $|f_3/f_6|$ is large, which shows that SSB only occurs to $S_3$.}
\label{4lobes}
\end{figure}

Specifically, we will study the shape of the energy $E$, and show that it cannot develop more than $n$ minima. This will be true if $E$ has at most $3$ minima in each angular direction when the simplex is embedded in $\mathbb{R}^2$%\footnote{In other words, each 2-simplex of the $n-1$ simplex has $\mathrm{D}_6$ (dihedral) symmetry.}
, as illustrated in Fig. \ref{4lobes}. Due to permutation symmetry, we can just pick any angular coordinate $\theta$ and write $E$ as 
\begin{equation}
E(R,\theta) = \sum_{k=0}^\infty a_{3k}(R) \cos(3k\theta).
\end{equation}
where $R$ is proportional to the distance from the axis of rotation of $\theta$:  see Subappendix \ref{barycentric} for details on how to find $R$ and $\theta$. Due to the $\mathrm{D}_6$ (dihedral) symmetry of each 2-dimensional simplex, only Fourier modes of order $3k$ exist. 

Any extremum of $E$ occurs when
\begin{equation}
\partial_\theta E=\sum_{k=1}^\infty 3k a_{3k}(R)\sin(3k\theta)=0.  \label{eqf2}
\end{equation}We wish to find the circumstances under which this equation \emph{cannot} have any other root other than the six trivial ones at $\theta = \pi n /3$, for $n=1,\ldots,6$. A sufficient condition for this is
\begin{equation}
\sum_{k=2}^\infty  k|a_{3k}(R)| < \frac{2}{\pi} a_3(R).  \label{eqg4}
\end{equation}To prove this, observe that, e.g. for $|\theta|<\pi/6$, we have $|\sin (3x)| > (2/\pi) 3|x|$, and $|\sin (3nx)| < 3n|x|$;  we then split up the sum Eq. (\ref{eqf2}) into the $n=1$ piece and the $n>1$ piece, and apply these bounds to each side.    By symmetry, an analogous argument holds for the other 5 regions of the circle $0\le \theta <2\pi$. 

Hence to show that the SSB pattern is $\mathrm{S}_n\rightarrow \mathrm{S}_{n-1}$, we just have to show that Eq. (\ref{eqg4}) is satisfied within some finite distance $r_*$ from the permutation symmetric point; it will turn out that $r_* \sim 10/\beta$, which implies that the spontaneously broken minima favor one choice until $\beta$ is large (at which point Eq. (\ref{Gzerotemp}) is valid).

\begin{figure}
\centering
\includegraphics[scale=0.99]{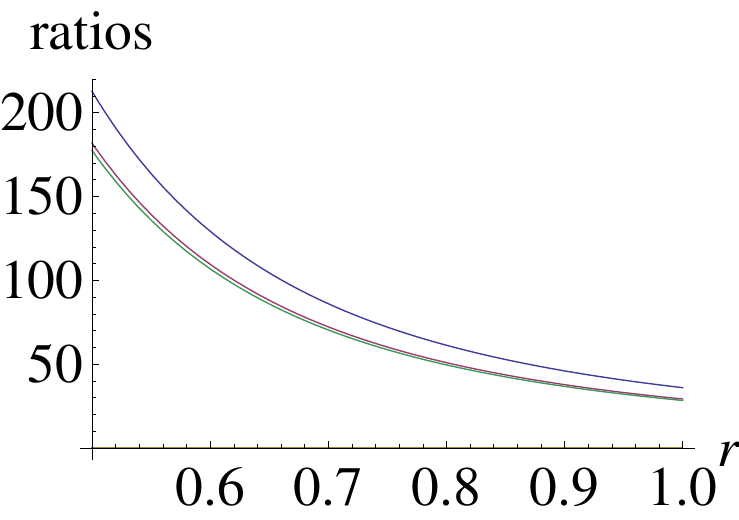}
\caption{Comaparison between the numerically computed $|a_{3k}|/|a_{3(k+1)}|$ for $k=1 $(blue), $k=2$ (red), and its theoretical prediction $|z_0|^{-3}$ (green), all at $\beta=4$. We see that the ratios are much greater than one, which suggests that higher harmonics leading to additional SSB are strongly suppressed. The numerically computed ratio of $|f_3/f_6|=c|z_0|^{-3}$, where $c$ differs from unity by less than $15\%$. The ratio between successive harmonics like $|f_6/f_9|$ is almost indistinguishable from the complex analytic result.  }
\label{aratio}\end{figure}

The decay rate of $a_k(R)$, the Fourier coefficients of $E(R,\theta)$, is controlled by the complex analytic properties of $E(R,z)$ where $z=\mathrm{e}^{\mathrm{i}\theta}$. From Eq. (\ref{decaykproof}) in Section \ref{fourieranalytic}, the coefficients decay like %\red{This is not supposed to be a lower or upper bound; it is a saturated limit asymptotically. For smaller k, deviations can be either positive or negative}
\begin{equation}
\frac{|a_{3k}(R)|}{|a_{3(k-1)}(R)|}\sim |z_0|^3
\label{decayk}
\end{equation}
where $z_0$ is the closest singularity of $E(R,z)$ from the unit circle such that $|z_0|<1$.  While Eq. (\ref{decayk}) only holds asymptotically in the large $k$ limit, we numerically find for the logistic models it is accurate to $\approx 15\%$ for $|a_6|/|a_3|$ and $\approx 2.5\%$ for $|a_9|/|a_6|$ as shown in Fig. \ref{aratio}. Henceforth, we shall keep track of the largest discrepancy via the parameter $c=\frac{|a_3||z_0|^3}{|a_6|}\approx 1$. 

$E$ differs from $G$ by a trivial quadratic factor from $G$, whose functional form of is given by Eq. (\ref{gfdd}) with $\alpha_k=\mathrm{e}^{\beta q_k}$, $\beta$ being the market heterogeneity. As explained previously, singularities only occur when denominator $e^{\beta q_i}-e^{\beta q_j}=0$ \emph{and} $q_i \ne q_j$.  This occurs when
\begin{equation}
q_i(R,\theta) - q_j(R,\theta) = \frac{2\pi \mathrm{i}p}{\beta},  \label{eqg5}
\end{equation}for $i\neq j$ and $p$ a non-zero integer. Expressions for $q_i$ and $q_j$ may be found using Eq. (\ref{eqg10}). Due to permutation symmetry, we are free to choose any pair $i,j$. It turns out that $i=n-1,j=n$ gives the cleanest computation:
\begin{eqnarray}
q_{n-1}-q_n&=&\frac{2(n-1)}{n}x_{n-1} \prod_{j=1}^{n-2} S_j \notag\\
&=& \frac{2r(n-1)}{n} \sqrt{\frac{n}{2(n-1)}} \prod_{j=1}^{n-2} \sin \varphi_j \notag\\
&=& \frac{2R(n-1)}{n} \sqrt{\frac{n}{2(n-1)}} \sin\theta \notag\\
& =&  \frac{2\pi \mathrm{i}p}{\beta}
\end{eqnarray}
where we have used $S_k^2=1-(n-k)^{-2}$ from lines $1$ to $2$, and identified $R\sin\theta$ with $r\prod_{j=1}^{n-2} \sin \varphi_j $ in line $3$ (see Section \ref{barycentric}). Analytically continuing $\theta$ to $z=\mathrm{e}^{\mathrm{i}\theta}$, the above reduces to  

%Let us fix $\varphi_j$ for $1 \le j \le n-3$, and set $\varphi_{n-2} = \theta$ to be our single variable.   Also let $Z \equiv \prod_{j=1}^{n-3} \sin \varphi_j$; note $|Z| \le 1$.   Then writing $\theta$ in terms of $z=\mathrm{e}^{\mathrm{i}\theta}$,
\begin{equation}
-\sqrt{\frac{2n}{n-1}} \frac{2\pi p}{\beta R} =  \left(z-\frac{1}{z}\right)
\end{equation}
whose root $z_0$ closest to the unit circle gives us the decay rate of $|a_k(R)|$. %An upper bound of $|z_0|$ can easily be seen to be %\red{Note: I have an extra factor of $2$ here}
%\begin{eqnarray}
%|z_0|&=&\sqrt{1+\frac{2n}{n-1} \left(\frac{\pi p}{\beta R}\right)^2}- \sqrt{\frac{2n}{n-1}} \frac{\pi p}{\beta R}\notag\\
%&\le& \frac{\beta R}{2\pi} \sqrt{\frac{n-1}{2n}}\notag\\
%&\le& \frac{\beta r}{2\pi} \sqrt{\frac{n-1}{2n}}.
%\end{eqnarray}

Combining Eqs. (\ref{decayk}) and (\ref{eqg4}), the condition for $\mathrm{S}_n\rightarrow \mathrm{S}_{n-1}$ is \begin{equation}
\frac{2c}{\pi }>\sum_{k=1}^\infty (k+1)|z_0|^{3k},
\end{equation}
which leads to
\begin{equation}|z_0|^{-3}>1+\frac{\pi}{2c}\left(1+\sqrt{1+\frac{2c}{\pi}}\right)\approx 4+3(1-c).
\label{z0ineq}\end{equation} 
With \begin{equation}
|z_0|=\sqrt{1+\frac{2n}{n-1} \left(\frac{\pi p}{\beta R}\right)^2}- \sqrt{\frac{2n}{n-1}} \frac{\pi p}{\beta R},
\end{equation}
we obtain
\begin{eqnarray}
\frac{R}{p}&=&\frac{2\pi}{\beta}\sqrt{\frac{2n}{n-1}}\frac{|z_0|}{1-|z_0|^2}\notag\\
&\lesssim & \frac{8+4(c-1)}{\beta}\sqrt{\frac{n}{n-1}}
%&\approx& \frac{16.21}{\beta}
%&\sim & \frac{2^{5/6}}{\sqrt[3]{\pi n}}+\frac{\sqrt{8}}{\pi n}%\frac{3\pi}{\sqrt{2}\beta}(2\pi n + 2+ n)
\end{eqnarray}
where we have also used Eq. (\ref{z0ineq}). This inequality must hold for all values of $p$ and angular variables; in particular, for $p=1$ and $R=r$ when the LHS is maximal. As previously mentioned, the error $|c-1|=\left||a_3||z_0|^3/|a_6|-1\right |\sim 0.1$, so we conclude that $E$ is broken as $\mathrm{S}_n\rightarrow \mathrm{S}_{n-1}$ for
\begin{equation}
r<r_*\approx \frac{8}{\beta}\sqrt{\frac{n}{n-1}}
\end{equation}

For $\beta\sim \mathrm{O}(1)$, $r_*>1$ and we have $\mathrm{S}_n\rightarrow \mathrm{S}_{n-1}$ for the entire configuration space. But we also know that the same SSB pattern occurs in the  $\beta\rightarrow \infty$. 
%Hence we have shown that permutation symmetry is indeed always at least $\mathrm{S}_{n-1}$ for generic logistic distributions with $n\lesssim 10$, and likely other unimodal distributions.

%When the small numerical deviations from Fig. \ref{aratio} has been taken into account, the coefficient of $8.394$ above will be changed by a comparably small percentage. This will not change the abovementioned conclusion for all values of $\beta$. 

%\red{Important to show a bit of numerics, since fourier result is only rigorous for large powers}

%A theorem from complex analysis states that (see the subsection to this appendix for a proof) if we have a function \begin{equation}
%f\left(\mathrm{e}^{\mathrm{i}\theta}\right) = \sum_n \lambda_n \cos(n\theta),
%\end{equation}and asymptotically at large $n$, $\lambda_n \sim \mathrm{e}^{-hn}$, with $h>0$, then this is equivalent to the statement that in the complex plane, the closest singularity to the unit circle ($|z|=1$) of $f(z)$ occurs at a point $z_0$, with $|z_0|=\mathrm{e}^{-h}$.    

%Now, we need to bound the Fourier components $a_{3k}(r)$.   Using Eq. (\ref{gfdd}) and the above theorem, we see that it suffices to study the locations of the solutions (with $q_i \ne q_j$) to \begin{equation}
%q_i - q_j = \frac{2\pi \mathrm{i}p}{\beta},  \label{eqg5}
%\end{equation}for $p$ a non-zero integer.   

\subsection{Barycentric coordinates for $\vec q$ in $\mathbb{R}^{n-1}$.}
\label{barycentric}
As we have done before in Eq. (\ref{baryeq}), we will want to find an embedding of $\vec q$ onto the $(n-1)$-dimensional simplex in $\mathbb{R}^{n-1}$. The most straightforward way to do this is to write \begin{equation}
\vec x = \sum_{k=1}^n q_k \vec b_k,
\end{equation}where $\vec x$ and $\vec b_k$ are $(n-1)$-dimensional (linearly dependent) basis vectors, normalized so that \begin{equation}
\vec b_k \cdot \vec b_j = \frac{n}{n-1}\delta_{jk} - \frac{1}{n-1}.
\end{equation} Geometrically, the $b_k$s give the positions of the  vertices of the simplex, and are at angle of $\arccos(-1/(n-1))$ from one another. We will wish to choose the point $\vec x = \vec 0$ corresponds to the permutation symmetric point. Intuitively, the $b_k$s should point towards the vertices of the simplex representing the various choices, and should be all   ion symmetric point. A basis consistent with all the above requirements is \begin{subequations}\begin{align}
\vec b_1 &= (1,0,\ldots,0), \\
\vec b_2 &= (C_1, S_1,\ldots,0), \\
\vec b_3 &= (C_1, S_1C_2, S_1S_2,\ldots,0), \\
&\vdots \notag \\
\vec b_{n-1} &= (C_1,S_1C_2,S_1S_2C_3,\ldots, S_1\cdots S_{n-2}), \\
\vec b_n &= (C_1,S_1C_2,S_1S_2C_3,\ldots, -S_1\cdots S_{n-2}),
\end{align}\end{subequations}with\begin{equation}
C_k = -\frac{1}{n-k}, \;\;\;\; S_k^2+C_k^2 = 1.
\end{equation}Note that, given the simplex constraint Eq. (\ref{simplexeq}), we can easily check that \begin{equation}
\vec q_k = \frac{1}{n}+ \frac{n-1}{n} \vec x \cdot \vec b_k.  \label{eqg10}
\end{equation}
and that 
\begin{equation}
|\vec x|^2 = \frac{n}{n-1}\sum_k^n\left(q_k-\frac{1}{n}\right)^2
\label{eqg11}
\end{equation}
For the purposes of this Appendix, we will also need to express $\vec x$ explicitly in terms of angles in orgin-centered spherical coordinates, analogously to the $n=3$ case: \begin{subequations}\begin{align}
x_1 &= r \cos \varphi_1, \\
x_2 &= r \sin\varphi_1 \cos\varphi_2, \\
&\vdots \notag \\
x_{n-2} &= r  \cos \varphi_k \prod_{k=1}^{n-3}\sin\varphi_k, \\
x_{n-1} &= r \prod_{k=1}^{n-2}\sin\varphi_k.
\end{align}\end{subequations}

\subsection{Proof of the Decay of Fourier Coefficients }
\label{fourieranalytic}

Here we prove that the fourier coefficients $a_k$ of a periodic function $E(z)$, where $z=\mathrm{e}^{\mathrm{i}\theta}$, decay asymptotically like 
\begin{equation} |a_k|\sim \lambda^k \label{decaykproof}\end{equation}
up to a proportionality factor, where $\lambda = |z_0| <1$, with $z_0$ is the closest singularity of $E(z)$ from the unit circle with $|z|_0<1$.  In particular, there is a constant $C$ such that $a_k <C\lambda ^k$. This result is well-known \cite{kohn1959,he2001}, but here we shall provide a simpler derivation suitable for our context.  Recall the definition that
\begin{equation} E(z)= \sum_{k\ge 0} \frac{a_k}{2} \left(z^k + \frac{1}{z^k}\right) \label{fourierseries}\end{equation}
Since $E(z)$ is analytic for $|z|>|z_0|$ within the unit circle, the above series must converge in that region.   As Eq. (\ref{decaykproof}) must hold for some value of $\lambda$ for this series to converge at all inside the unit circle, let us assume this holds and determine the required value of $\lambda$.   When $\lambda <|z|<1$: \begin{equation}
E(z) < \sum_{k\ge 0} \frac{|a_k|}{|z|^k} < C \sum_{k\ge 0} \left|\frac{\lambda}{z}\right|^k < \infty
\end{equation}In addition, $E(z)$ fails to be analytic at $z_0$, so the above series must diverge when $|z|=\lambda$.   Evidently, $|z_0|=\lambda$. This implies that $|a_k|$ must asymptotically decay as $|z_0|^k$, proving Eq. (\ref{decaykproof}).

%Alternatively, we can proof it by looking at the contour version of the fourier coefficients $a_k=\frac{1}{2\pi i}\oint_{|z|=1} E(z)z^k\frac{dz}{z}$, and then deforming the contour to a circle of radius infinitesimally larger than $|z_0|$.

\end{document}